\documentclass[letterpaper,twocolumn,showpacs,10pt]{revtex4-1}
\pdfoutput=1
\usepackage{graphicx,amsmath,mathrsfs}
\usepackage{amssymb}
\usepackage{amsthm}
\usepackage{bm}
\usepackage{float}
\usepackage{epsfig}

\usepackage{color}

\def\bc{\begin{center}}
\def\ec{\end{center}}

\renewcommand{\b}{\beta}

\renewcommand {\ec}{\eta_{\gamma}}

\newcommand{\comment}[1]{}

\begin{document}
\title{Exact holographic mapping in free fermion systems}
\author{Ching Hua Lee}
\email{calvin-lee@ihpc.a-star.edu.sg}
\affiliation{Institute of High Performance Computing, 138632, Singapore}
\author{Xiao-Liang Qi}
\affiliation{Department of Physics, Stanford University, Stanford, CA 94305, USA}

\date{\today}
\begin{abstract}
In this paper, we perform a detailed analysis of the Exact Holographic Mapping first introduced in arXiv:1309.6282, which was proposed as an explicit example of holographic duality between quantum many-body systems and gravitational theories. We obtain analytic results for free fermion systems that not only confirm previous numerical results, but also elucidate the exact relationships between the various physical properties of the bulk and boundary systems. These analytic results allow us to study the asymptotic properties that are difficult to probe numerically, such as the near-horizon regime of the black hole geometry.  We shall also explore a few interesting but hitherto unexplored bulk geometries, such as that corresponding to a boundary critical fermion with nontrivial dynamical critical exponent. Our analytic framework also allows us to study the holographic mapping of some of these boundary theories in dimensions 2+1 or higher.
\end{abstract}
\maketitle

\tableofcontents

\section{Introduction}
In the recent years, holographic duality, also known as the Anti-de-Sitter space/Conformal Field Theory (AdS/CFT) correspondence\cite{maldacena1998,witten1998,gubser1998}, has attracted tremendous research interest in both high energy and condensed matter physics. This correspondence is defined as a duality between a $D+1$-dimensional field theory on a fixed background geometry and a $D+2$-dimensional quantum gravity theory. 
The best understood example of holographic duality is the correspondence between $4$-dimensional super-Yang-Mills theory and $5$-dimensional supergravity. There, the large-$N$ limit of the super-Yang-Mills theory corresponds to the classical limit of the dual gravity theory, which provides a helpful description of strongly coupled gauge theories. What makes holographic duality particularly interesting is its generality. When the boundary theory is not a conformal field theory, a dual theory with a different space-time geometry may still be well-defined.\cite{witten1998b} Physically, holographic duality can be understood as a generalization of the renormalization group (RG) flow of the boundary theory\cite{akhmedov1998,boer2000,skenderis2002}, where bulk gravitational dynamics generalize the RG flow equations and the emergent dimension perpendicular to the boundary has the physical interpretation of energy scale\cite{heemskerk2011,lee2010}. 
Indeed, holographic duality has been applied to condensed matter physics as a new tool to characterize strongly correlated systems\cite{hartnoll2009,horowitz2009,mcgreevy2010,sachdev2012}.

More recently, holographic duality has been proposed to be related to another approach developed in condensed matter physics, namely tensor networks\cite{swingle2012,swingle2012b,evenbly2011,nozaki2012,hartman2013,czech2014,miyaji2014,miyaji2015,pastawski2015}. 
In its most general form, tensor networks refer to a description of many-body wavefunctions and operators (i.e. linear maps) by contracting tensors defined on vertices of a graph.\cite{white1992,klumper1993,verstraete2004,vidal2007,vidal2008,gu2009} 
More specifically, the tensor network state proposed to be related to holographic duality is the multiscale entanglement renormalization ansatz (MERA)\cite{vidal2007,vidal2008}, which is defined on a graph with hyperbolic structure, with external indices (corresponding to the physical degrees of freedom) at the boundary and internal indices contracted in the bulk. An important feature of states described by tensor networks is that the entanglement entropy of a given region is bounded by the number of links between the region and its complement. This property motivated its relation to holographic duality\cite{swingle2012}, where the entanglement entropy of a given region is determined by the area of the minimal surface bounding it, in accordance to the Ryu-Takayanagi formula\cite{ryu2006}. 

There are many open questions in the proposed tensor network interpretation of the holographic duality. One important question is how to describe space-time geometry rather than spatial geometry. Another (related) question is how to understand excitations (quantum fields) living in the bulk. Motivated by these questions, one of us\cite{qi2013} proposed a tensor network which defines not a many-body state but a unitary mapping between the boundary and bulk systems, known as the exact holographic mapping （EHM）. The EHM is a tensor network very similar to MERA, except that it is a one-to-one unitary mapping between boundary and bulk degrees of freedom. Each boundary state $|\psi\rangle$ is mapped to a bulk state $|\tilde{\psi}\rangle=M|\psi\rangle$, and each boundary operator $O$ is mapped to a bulk operator $\tilde{O}=MOM^{-1}$. Physically, the EHM is a ``lossless" version of real space renormalization group. Denoting a site in the bulk as ${\bf x}$, a local operator at that site $\tilde{O}_{{\bf x}}$ is dual to a generically nonlocal operator on the boundary $O_{\bf x}=M^{-1}\tilde{O}_{{\bf x}}M$. Different bulk sites ${\bf x}$ correspond to operators $O_{\bf x}$ on the boundary with different energy scales and different center-of-mass locations of their support. Once a mapping $M$ is chosen, bulk correlation functions can in principle be calculated. Motivated by the general principle of relativity, the bulk geometry was proposed to be determined by the bulk correlation functions. More specifically, the distance between two points was proposed to depend logarithmically on the connected two point correlation functions. Compared to previous tensor network proposals, the EHM is different in two aspects: i) The bulk geometry is not determined by the structure of the tensor network but by the correlation structure of the bulk state; ii) The bulk geometry can be studied in both the spatial and temporal direction by studying the bulk correlation functions. In Ref. \onlinecite{qi2013}, an explicit choice of the mapping $M$ for $(1+1)$-dim lattice fermions was proposed, and the consequent dual geometries corresponding to different boundary states were studied. They included the ground state of massless and massive fermions, the nonzero temperature thermal ensemble of massless fermions, and a thermal double state which is a purification of the thermal ensemble. Dynamics after a quantum quench was also studied in the thermal double system, motivated by a comparison with geometrical properties of a two-sided black hole space-time\cite{hartman2013}.

The results in Ref. \onlinecite{qi2013} for the abovementioned free fermion systems were obtained numerically. This limits the extent of analysis, due especially to the exponential growth of boundary system size. To have a well-defined bulk geometry with $N$ layers in the emergent direction of the bulk perpendicular to the boundary, the boundary system has to have $2^N$ sites. In this paper, we shall obtain analytic results on the free fermion EHM, which will enable us to rigorously determine asymptotic properties of the dual geometry, and also to discuss more general boundary systems. For instance, the existence of a black hole horizon in the geometry dual to a nonzero temperature state at the boundary can be studied more explicitly from the \emph{asympotic} infrared behavior of correlation functions in both spatial and temporal directions. In addition to reproducing the results of Ref. \onlinecite{qi2013} analytically, we shall also explore a few other interesting emergent bulk geometries, such as that corresponding to a critical fermion with nontrivial dynamic critical exponent. Our analytic framework also allows us to generalize the EHM to boundary theories with dimension $2+1$ or higher\footnote{The higher dimensional generalization of EHM is also independently investigated by Xueda Wen, Gil Y. Cho and Shinsei Ryu.}, in which case the analytic approach is more essential due to the increasing difficulty of numerical calculations\footnote{In $(1+1)$-dimension, at least $2^{15}$ sites are needed for analyzing the dual geometry with reasonable precision. In $(2+1)$-dimensions the same number of layers in the bulk will require $2^{30}$ sites.}. An added advantage of an analytic approach is that it allows one to identify properties of the bulk geometry that are insensitive to details of the choice of the mapping which thus reflects intrinsic properties of the boundary state. 

This paper is structured as follows. In Section II, we first review the EHM construction by describing its general principles and the definition of bulk geometry. These ideas will be elaborated in Section III for free lattice fermions, where an explicit Haar wavelet representation of the EHM will be presented. In Section IV, we provide detailed descriptions of the asymptotic correlator behavior and corresponding bulk geometries for the prototypical $1+1$-dim Dirac model at various combinations of zero and nonzero temperature and mass. These developments will be further extended to higher dimensions and generic energy dispersions in Section V, where we discuss the emergence of interesting geometries like anisotropic black hole horizons with nontrivial topology.

\section{Review of the Exact Holographic Mapping}\label{sec:review}

In this section, we shall review the motivation and construction of the EHM proposed in Ref. \onlinecite{qi2013} in a formalism that will be helpful for the later part of this paper. We will also include some new insights that are not discussed in the original proposal. The EHM approach is defined by the following two principles:
\begin{enumerate}
\item The bulk theory and boundary theory are defined in the same Hilbert space. The bulk local operators are determined by a unitary mapping acting on the boundary local operators.
\item The bulk geometry is determined by physical correlation functions. More specifically, the distance between two space-time points ${\bf x}, {\bf y}$ in the bulk is determined by the connected correlation functions between the two points.
\end{enumerate}
Although the abovementioned unitary transformation can be very generic in principle, the types of transformations that are relevant for holographic duality are those which are physically analogous to the renormalization group\cite{wilson1974,wilson1975}. The bulk operators at different locations should represent boundary degrees of freedom with different energy scales. The key difference from the conventional RG approach is that the high energy degrees of freedom are spatially separated from low energy ones, instead of being integrated out. This enables us to concretely answer many new questions, such as how the high and low energy degrees of freedom (DOFs) are entangled/correlated. In the following, we will elaborate on the two abovementioned principles in the context of free fermion systems, and discuss the transformation of free fermion Hamiltonians under EHM.

\subsection{General construction of EHM}

\begin{figure}[H]
\includegraphics[scale=.28]{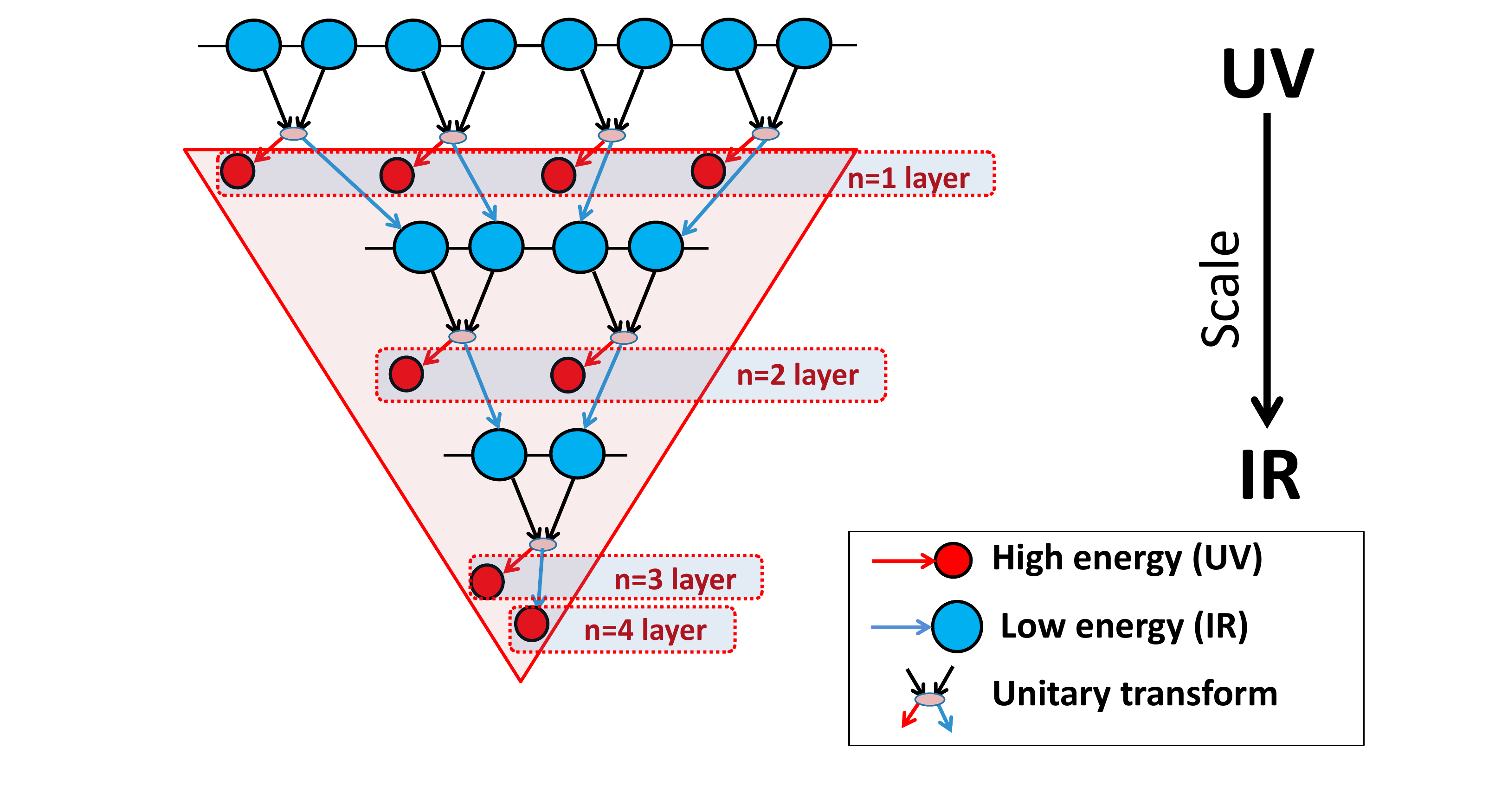}
\caption{(Color Online) Illustration of the EHM on $2^3=8$ sites. At each iteration, two auxiliary sites (blue) is fed into a unitary transform that produces a UV (red) DOF that defines a bulk site, and a IR (blue) DOF that becomes the auxiliary site for the next iteration. The bulk sites are arranged in a tree-like structure (red triangle) with $4$ layers, inclusive of the last (lowest energy) IR that forms the last "layer". }
\label{ehmtree}
\end{figure}

The Exact Holographic Mapping is a unitary transformation defined by a tensor network or, equivalently, a quantum circuit consisting of local unitary operators. As proposed in Ref. \onlinecite{qi2013}, a simple construction of the EHM is given by a tree-shaped tensor network depicted in Fig. \ref{ehmtree}, where bulk (red) sites at the same level belong to the same \textsc{\char13}layer\textsc{\char13}. To construct it, we first take a $D+1$-dimensional boundary system to be the zeroth bulk layer with $L^D=2^{ND}$ sites. To construct the first bulk layer, one performs a unitary transform $U$ on every set of $2^D$ adjacent sites such that the UV and IR (high and low momentum, assuming a monotonic energy dispersion) degrees of freedom are separated out. For $D=1$, this can be written as
\begin{equation}
U_{12}|\psi_1\psi_2\rangle =\sum_{\alpha,\beta} U^{\alpha\beta}_{\psi_1\psi_2}|\alpha\rangle_{IR} |\beta\rangle_{UV}
\label{unitary}
\end{equation}
where $U_{12}$ only acts on states $\psi_1,\psi_2$ on sites $1$ and $2$ respectively, and $|\alpha\rangle $ and $|\beta\rangle$ capture the higher and lower momentum (shorter and longer scale) degrees of freedom respectively. The construction of these $|\alpha\rangle $ and $|\beta\rangle$ states will be shown in detail in the next section. The full transformation on the zeroth layer is given by
\begin{equation}
U= U_{12}\otimes U_{34}\otimes ...\otimes U_{2^N-1,2^N},
\label{U}
\end{equation}
which is a unitary transform on the Hilbert space of the whole layer. For $D>1$ dimensions, $U$ will be given by the direct product of $D$ copies of the expression in Eq. \ref{U}.

We construct the first bulk layer from the component $|\beta\rangle=|\beta\rangle_{UV}$ in Eq. \ref{unitary}, which has the UV half of the degrees of freedom in the original layer. The other lower energy half $|\alpha\rangle$, which we shall call the \emph{auxillary sites} in deference to Ref. \onlinecite{qi2013}, are fed into another copy of $U$ with half the number of sites. This process is iterated for $N$ times, each time producing a new layer in the bulk that has $1/2^D$ the number of sites as the preceding layer, until only one site is left. The resultant (bulk) tree\footnote{In the continuum limit, the bulk system is topologically half the suspension (cone) of the boundary system, i.e. the latter with successively smaller copies of itself connected in a prism-like manner. Loosely speaking, the bulk system can be visualized as the solid 'interior' of the boundary manifold.} is unitary equivalent to the original (boundary) system, and is illustrated in Fig. \ref{ehmtree}.

\subsection{Emergent bulk geometry through boundary correlators}

The key motivation behind the EHM approach is to uncover the relationship between space-time geometry and the quantum entanglement properties of a quantum many-body system. The unitary mapping defined by the tensor network defines a new direct-product decomposition of the Hilbert space , and is chosen to make physical correlation functions more local in this new basis. To be more precise, we assume that the two-point connected correlation functions in the bulk \emph{always} decay exponentially, according to the geodesic distance of certain emergent geometry:
\begin{eqnarray}
C_{({\bf x}_1,t_1)({\bf x}_2,t_2)}&\equiv&\left\langle O_{\bf x_1}(t_1)O_{\bf x_2}(t_2)\right\rangle-\left\langle O_{\bf x_1}(t_1)\right\rangle\left\langle O_{\bf x_2}(t_2)\right\rangle\nonumber\\
&\propto &\exp\left(-d_{({\bf x}_1,t_1),({\bf x}_2,t_2)}/\xi\right)
\end{eqnarray}
This assumption can conversely be used as a definition of the distance\cite{qi2013}:
\begin{equation}
d_{(\bold x_1,t_1),(\bold x_2,t_2)}= -\xi \log \frac{C_{({\bf x}_1,t_1)({\bf x}_2,t_2)}}{C_0}
\label{d}
\end{equation}
where $C_0$ and $\xi$ control the overall offset and scaling respectively. $\xi$ can be physically interpreted as the inverse mass of the emergent bulk theory, which may depend slightly on how we perform the EHM. The logarithmic dependence is physically motivated by the observation that for a massive system, $d_{(\bold x_1,t_1),(\bold x_2,t_2)}$ should recover the Euclidean distance in the original system.

In this work, we shall for simplicity focus on systems that are translationally-invariant in space and time, and study only correlators with purely space or time intervals, i.e. $\Delta t=t_2-t_1=0$ or $\Delta x =x _2-x_1=0$.

In the former case with purely spatial interval, all two-point connected correlators are bounded above\cite{wolf2008} by the \emph{mutual information}
\begin{equation}
I_{\bold x \bold y}=S_{\bold x} + S_{\bold y} - S_{\bold x \bold y}
\label{mutual}
\end{equation}
where $S_{\bold x }$ and $S_{\bold x \bold y}$ are the entanglement entropies (EE) of a single site and two sites respectively. Roughly speaking, the mutual information between two sites measures how much the entanglement entropy of two sites will be reduced if the correlation between the two sites are known. Hence a basis-independent definition of the spatial geometry is given by the mutual information:
\begin{equation}
d_{\Delta \bold x}
=-\xi \log \frac{I_{\bold x \bold y}}{I_0}
\label{dx}
\end{equation}
where $\Delta \bold x=|\bold x -\bold y|$ and $I_0$ is a reference value for the mutual information. One reasonable choice of $I_0$ is $I_0=2\aleph\log 2 $, which is the maximal mutual information between two sites, each with $\aleph$ internal DOFs (spins, bands, etc). This bound is saturated in the (hypothetical) situation when $S_{\bold x \bold y}=0$ but $S_{\bold x}=S_{\bold y}=\aleph\log 2$, i.e. when the DOFs of the two sites are maximally entangled with each other but not with those of the other sites. A more detailed explanation for the mutual information is given in Appendix \ref{app:mutual}.

In the latter case with purely temporal interval, we specialize Eq. \ref{d} to
\begin{equation}
d_{\tau}= -\xi_\tau \log \frac{C(\tau)}{ C(0)}
\label{dt}
\end{equation}
where $C(\tau)$ represents chosen component/s of $\langle O_{\bold x}(0) O_{\bold x}(\tau)\rangle_{bulk}$, $\tau$ being the imaginary (Wick-rotated) time interval $\tau=-i\Delta t$. The imaginary time direction is preferred over real time as the latter typically exhibits oscillatory behavior that makes an asymptotic comparison difficult. Further discussion on the relationship between the real and imaginary time correlators will be deferred to Appendix \ref{sec:time}. Note that unlike the case with spatial intervals, there is no known operator that yields the upper bound of correlators across temporal intervals.

Henceforth, we shall use Eq. \ref{dx} and \ref{dt} as the expressions for the distance between two points in the bulk, and compare them with the geodesics of classical geometries.

\section{Exact holographic mapping for free lattice fermions}

We now specialize the above developments to free lattice fermions, for which the correlators and mutual information possess nice analytic behavior, at least asymptotically. First, we recall the following well-known result for the entanglement entropy of free fermions\cite{peschel2002,klich2006, lee2014exact}:
\begin{equation}
S_X=-\text{Tr }(C_X\log C_X + (\mathbb{I}-C_X)\log(\mathbb{I}-C_X))
\label{Sx}
\end{equation}
where $S_X=-\text{Tr }(\rho_X\log\rho_X)$ is the entanglement entropy for the region $X$, and $C_X$ is the projector (correlator) onto region $X$. With the help of Eq. \ref{Sx}, it is shown in Appendix \ref{app:mutualcorr} that the Mutual Information is approximately
\begin{eqnarray}
I_{\bold x \bold y  }&=&S_\bold x +S_\bold y  -S_{\bold x \bold y  }\notag\\
&\approx& \frac{1}{2}\text{Tr }\left[C_{\bold x -\bold y  }\frac{1}{C_\bold y  (\mathbb{I}-C_\bold y  )}C_{\bold y  -\bold x }+(\bold x \leftrightarrow \bold y  )\right]\notag\\
&\sim& \text{Tr } [C^\dagger_{\bold y  -\bold x }C_{\bold y  -\bold x }]
\label{Ixy00}
\end{eqnarray}
where $C_\bold x ,C_\bold y  $ are the single-particle onsite correlators, and $C_{\bold x -\bold y  }$ is the \emph{single-particle} propagator between the \emph{two different sites} $\bold x $ and $\bold y  $. This result is completely general, and implies that
\begin{equation}
d_{\Delta \bold x }=-\xi_{\Delta \bold x }\log \frac{I_{\bold x \bold y  }}{I_0}\sim 2\xi_{\Delta \bold x } \log \text{Tr } C_{\bold y  -\bold x }
\end{equation}
in the limit of large spatial separation $|\bold x -\bold y  |$. That $C_\bold x ,C_\bold y  $ drops out is hardly surprising, as they each depend only on one site, and have no knowledge about their separation. Indeed, most of the information transfer in the asymptotic limit is dominated by the single-particle propagator.

We also define the temporal distance via
\begin{equation}
d_{\tau}= -\xi_\tau \log \frac{\text{Tr } C(\tau)}{\text{Tr } C(0)}
\label{dt2}
\end{equation}
where a trace of the fermion states have been taken. This is the simplest possible basis-independent combination of the components of $C(\tau)$.

In the next two subsections, we shall introduce prototypical fermionic models as the boundary systems in $1+1$ and higher dimensions, and show how their corresponding bulk distances and hence geometries can be computed via suitable holographic unitary mappings.

\subsection{EHM for (1+1)-dimensional lattice Dirac fermions}
\label{subsec:kspace}

The (1+1)-dimensional lattice Dirac model is among the simplest models with a single critical point. In this subsection, we will summarize the explicit construction of the EHM for this system. Its simplicity allows us to study its multitude of entanglement and geometric properties analytically with minimal complication.

The $(1+1)$-dim Dirac hamiltonian is a 2-band hamiltonian given by
\begin{equation}
H_{Dirac}(k)=v_F[\sin k \sigma_1 + M(m+1-\cos k)\sigma_2]
\label{dirac1}
\end{equation}
where $\sigma_1,\sigma_2$ are the Pauli matrices and $v_F$, the Fermi velocity, controls the overall scale of the dispersion. $M$ is controls the relative weight between the $\sin k$ and $m+1-\cos k$ terms, and will be set to unity here. A discussion for generic $M$ will be given in Appendix \ref{genericdirac}. When $m=0$ or $\pm 2$, its gap closes at $k=0$ and it becomes critical with two crossing bands with linear dispersion. To explore or ''zoom into'' the low energy (IR) degrees of freedom (DOFs), we utilize a unitary transform that maps states $|\psi_1^{s_1}\rangle,|\psi_2^{s_2}\rangle$ on neighboring sites into symmetric (low energy) and antisymmetric (high energy) linear combinations $\frac1{\sqrt{2}}\left(|\psi_1^{s_1}\rangle\pm|\psi_2^{s_2}\right)$. Note that the unitary transform does not rotate the spin labels $s_1,s_2$, which we shall suppress in the following. In matrix form, the unitary transform is written as
\begin{equation}
U_{12}=\frac{1}{\sqrt{2}} \left(\begin{matrix}
 & 1 & 1 \\
 & 1 & -1\\
\end{matrix}\right)
\label{haarU}
\end{equation}
The symmetric combination has a Fourier peak at $k=0$, which is exactly the gapless point of the critical ($m=0$) Dirac model. The discerning reader will notice that $U_{12}$ is nothing other than the defining expression for the Haar transform. Indeed, the construction of the EHM basis is mathematically identical to performing a wavelet decomposition\cite{meyer1989,daubechies1992,strang1996}. A systematic study of all possible wavelet descriptions of the EHM will be deferred to future work, since for this work we will be primarily concerned about the behavior of the bulk geometries due to qualitatively different boundary systems, not the details of the wavelet mapping. The transform given by Eq. \ref{haarU} possess the virtue of simplicity and, most importantly, fixes the archetypal Dirac Hamiltonian, a property we shall prove in the next subsection.

\begin{figure}[H]
\includegraphics[scale=.95]{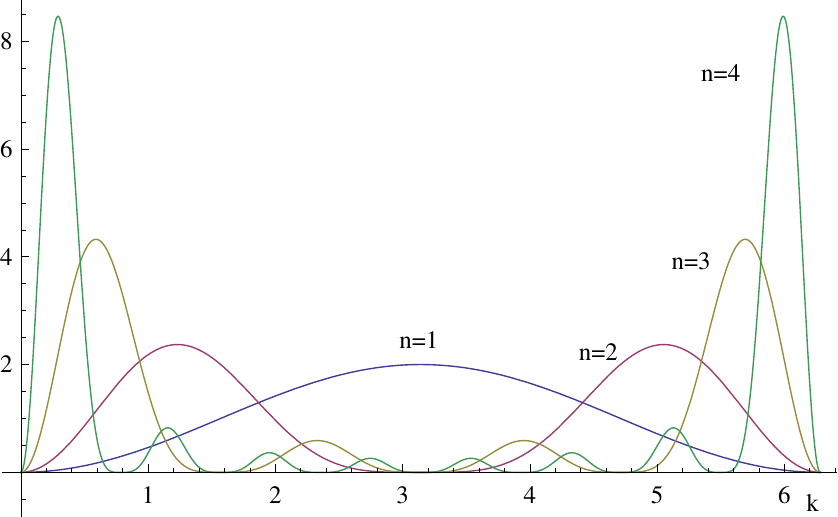}
\caption{(Color Online) Plot of the normalized spectral weight $|W_n(e^{ik})|^2$ of the bulk basis for $n=1,2,3,4$. We see that as $n$ increases, the dominant spectral peak approaches the unique 
IR point at $k=0$ (and its periodic image at $k=2\pi$)
exponentially viz. $k_0\approx \frac{2\pi}{2^n}\rightarrow 0$. Also, it becomes narrower since its spectral width also goes like $\sim 2^{-n}$. $W_1$ is peaked at the highest momentum $k=\pi$, attesting to the fact that it contains all the UV DOFs.   } \label{wavelets}
\end{figure}

More insight into the EHM can be gleaned in momentum space, where one can directly see how the Hilbert space is decomposed into layers with different momentum spectral distributions. Fourier transforming the action of Eq. \ref{haarU} on the single particle states, we obtain $|\alpha_k\rangle = \sum_{2k} C(e^{ik})|\psi_{2k}\rangle$ and $|\beta_k\rangle = \sum_{2k} D(e^{ik})|\psi_{2k}\rangle$ for the auxiliary and bulk states respectively, where $|\psi_k\rangle$ is the periodic part of the Bloch state and
\begin{equation}C(e^{ik})=\frac{1}{\sqrt{2}}\left(1+e^{ik}\right),\label{C}\end{equation}
\begin{equation}D(e^{ik})=\frac{1}{\sqrt{2}}\left(1-e^{ik}\right)\label{D}\end{equation}
We shall call $C,D$ the IR and UV (low energy and high energy) projectors. Physically, they represent the spectral weight projected to the auxiliary and bulk DOFs at each iteration. Through these iterations, we obtain successive basis projectors for each bulk layer that are increasingly sharply peaked in the IR. To understand this, note that the basis projector of the $n^{th}$ layer $W_n(z)=W_n(e^{ik})$ is obtained from $n-1$ consecutive IR outputs $|\alpha\rangle$ and one final UV output $|\beta\rangle$. Hence the first bulk layer should contain the DOFs projected from the UV projector $D(e^{ik})$, while the second layer should contain an IR projector $C(e^{ik})$ followed by an UV projector $D(e^{2ik})$ that peaks at half the momentum. This reasoning generalizes to all the $N$ layers, so the normalized projector for the $n^{th}$ layer is given in momentum space by (writing $z=e^{ik}$)
\begin{eqnarray}
W_n(z)&=&\frac{1}{\sqrt{2 \pi}}D\left(z^{2^{n-1}}\right)\prod_{j=1}^{n-1}C\left(z^{2^{j-1}}\right)\notag\\
&=&\frac{1}{\sqrt{2 \pi}}\frac{1-z^{2^n}}{\sqrt{2^n}} \prod_{j=1}^{n-1}\left(1+z^{2^{j-1}}\right)\notag\\
&=&\frac{1}{\sqrt{2 \pi}}\frac{1}{\sqrt{2^n}}  \frac{\left(1-z^{2^{n-1}}\right)^2}{1-z}.\notag\\
\label{Wn}
\end{eqnarray}
$W_n(e^{ik})$ contains a series of peaks interspersed by valleys at $e^{i2^{n-1}k}=1$. The dominant peaks occur at $k=\pm k_0\approx \frac{2\pi}{2^n}$ where the denominator is most singular, as shown in Fig. \ref{wavelets}, and has magnitude $|W_n(e^{ik_0})|=\sqrt{\frac{2^{n+1}}{\pi^3}}$. This means that as $n$ increases, the spectral weight of the $n^{th}$ bulk layer exponentially approach the IR point at $k=0$. One can further show that the $W_n$'s form a complete an orthonormal basis, i.e. $\int_{-\pi}^{\pi} W_m^*(e^{ik})W_n(e^{ik}) dk = \frac1{2\pi i}\oint_{|z|=1}W^*_m(z^{-1})W_n(z)\frac{dz}{z}=\delta_{mn}$, where the conjugation symbol in $W^*$ denotes that only the coefficients of $W(z)$, not the argument $z$, are complex conjugated. Indeed, that the $W_n$'s are orthonormal with peaks $k_0\sim 2^{-n}$ is testimony to the fact that the EHM is a unitary mapping that separates the momentum (or energy) scale.

Note that the auxiliary projector, i.e. projection to auxiliary sites with the IR (low energy) half of the DOF, is just the orthogonal complement of $W_n(z)$ in Eq. \ref{Wn}: It is given by $\frac{1}{\sqrt{2\pi}}\prod_{j=1}^n C(z^{2^{j-1}})$, comprising IR projectors $C(z)$ only.

From Fig. \ref{wavelets}, we see that the bulk basis from Eq. \ref{Wn} become more effective in separating different momentum (and hence energy) scales as $n$ increases, since they become more sharply peaked. This property is due to the recursive nature of the definition Eq. \ref{Wn}, which also implies that physical quantities, i.e. correlators \emph{must scale universally with $n$} in the IR limit. However, departures from universal scaling may occur at small $n$ (UV regime) due to the non-universal high energy characteristics of the boundary system.

Note that the abovementioned unitary EHM transform successively ''zooms into'' the low energy DOFs of any model with a critical point at $k=0$, and not just that of the Dirac model.


Having defined the mapping explicitly, we now write down the explicit expression of the bulk correlators. Previously, we have seen how the bulk distance can be expressed in terms of the bulk correlators $C_x, C_{x-y}$ and $C(\tau)$, which are the onsite, spatial and temporal propagators respectively. 
In the free fermion system discussed here, the bulk correlators are determined by the boundary two-point correlators
\begin{equation}
G_\bold q(\tau)=e^{\tau(H(\bold q)-\mu )}(\mathbb{I}+e^{\beta(H(\bold q)-\mu)})^{-1}
\label{gq0}
\end{equation}
where $H(\bold q)$ is the boundary single particle Hamiltonian matrix, $\mu$ the chemical potential and $T=1/\beta$ the temperature. Explicit expressions for $G_\bold q(\tau)$ as well as their resultant mutual information $I_{\bold x\bold y}$ are derived in Appendix \ref{app:twoband} For cases with particle-hole symmetry, which includes the Dirac model. In the zero temperature limit, $G_\bold q(0)$ reduces to a projector onto the occupied bands below the chemical potential. In the extremely high temperature limit, it becomes nearly the identity operator, which just means that almost every state is equally accessible. In the remainder of the paper, we will always focus on imaginary time correlator unless otherwise stated, since it is still unclear how to define time-drection distance from the rapidly oscillating real time correlators (as discussed further in Appendix \ref{sec:time}).

The bulk correlators are most easily expressed as a sum in momentum space, since we have already found projections to the various bulk layers in terms of the spectral weight. Taking the thermodynamic limit where the boundary system is infinitely large, i.e. $N\rightarrow \infty$, the sum over momenta can be replaced with an integral. The bulk correlator between two bulk points $(x_1,n_1,0)$ and $(x_2,n_2,\tau)$ is given by
\begin{eqnarray}
&&C(n_1,n_2, \Delta (2^n  x),\tau)\notag\\&=&\sum_{  q} W^*_{n_1}( e^{i q})W_{n_2}(  e^{iq})e^{i    q  \Delta (2^n  x) }G_{  q }(\tau)
\label{bulkcorrelator1d}
\end{eqnarray}
where $\Delta (2^n x)=2^{n_2}x_2-2^{n_1}x_1$ is the bulk angular interval and $G_{ q}(\tau)$ is the (matrix-valued) $(1+1)$-dim boundary correlator. The indices $n_1,n_2$ specify the coordinates ('layers') in the emergent 'radial' momentum-scale direction in the bulk. There is no translational symmetry in this momentum(or energy)-scale direction, unlike the original spatial and temporal directions.

We see that $C(n_1,n_2, \Delta   x,\tau)$ is just a Fourier transform of the boundary correlator $G_{  q}$ weighted by the spectral contributions $W_{n_1}^{*}W_{n_2}$ of the respective bulk layers. One important observation is that the spatial coordinate in the exponential factor is $\Delta (2^n   x)$, not $\Delta   x$. This is because the correlation comes from the holographic projection of the bulk sites onto the boundary, which depends on the angle subtended by the bulk displacement: In the simplest case of a circular boundary, the angle subtended by $\Delta x$ at the $n^{th}$ layer is $\Delta \theta = \frac{2\pi \Delta x}{2^{N-n}}= \frac{2\pi}{L} (2^n\Delta x)$. The bulk correlator still possesses translation invariance, but of $\Delta(2^n  x)$, the angular interval \emph{projected on to the boundary}, not of the bulk interval $\Delta x$ itself. Mathematically, we find that the $2^n$ rescaling of angular distance is also required for the orthogonality of the basis $\{W_{n}(  q)e^{i 2^n   q \cdot  x }\}$, $x=1,2,..., \frac{L}{2^n}$.

\subsection{Transformation of the Hamiltonian under the EHM and its fixed points}
The EHM is an exact version of renormalization group (RG) transformation. In each step, the DOFs of the system are split into the high energy (UV) and the low energy (IR) parts, with the procedure iterated on the low energy part. If we write the Hamiltonian in the new basis after an EHM step and ignore the coupling between the IR and UV degrees of freedom, we can write down a "low energy effective Hamiltonian" of the IR states. This resembles the renormalization group flow of the effective Hamiltonian in ordinary RG. For a given choice of the EHM tensor network, there are certain boundary systems for which the IR Hamiltonian is at an RG fixed point. In the following, we will show that the massless Dirac Hamiltonians are RG fixed points of the EHM transformation we defined earlier.

We start by writing down the effective IR Hamiltonian in the momentum basis. In $D=1$ spatial dimensions, the EHM for each iteration is given by the change of basis in Eqs. \ref{C} and \ref{D}, so the single particle Hamiltonian matrix $h^n$ of the $n^{th}$ layer is transformed according to
 \begin{equation} h^{n+1}(e^{ik/2}) = \left[V^\dagger \left(\begin{matrix}
 & h^n(e^{ik/2}) & 0 \\
 & 0 & h^n(e^{i(k/2+\pi)}) \\
\end{matrix}\right)V \right]_{11}
\label{RG0}
\end{equation}
with the two components of the matrix representing the IR and UV DOFs.  Here $V(w)=\frac{1}{\sqrt{2}}\left(\begin{matrix}
 & C(w) & D(w) \\
 & C(-w) & D(-w) \\
\end{matrix}\right)=\frac{1}{2}\left(\begin{matrix}
 & 1+w & 1-w \\
 & 1-w & 1+w \\
\end{matrix}\right)$, where $w=e^{ik/2}$. To obtain the Hamiltonian at the $(n+1)^{th}$ layer, one projects onto the upper left or IR component of the RHS. With $D>1$ spatial dimensions, $V$ will be given by $V(w_1)\otimes ... \otimes V(w_D)$. If the Hamiltonian is to remain invariant under the RG, the $11$ (IR) component in Eq. \ref{RG0} gives $2h^{n+1}(w)=2\lambda h^n(w^2)$ or, in detail:
\begin{eqnarray} 
&&2\lambda h^n(w^2)\notag\\
&=&h^n(w)C(w)C^*(w^{-1}) + h^n(-w) C(-w)C^*(-w^{-1}) \notag\\
&=& (h^n(w)+h^n(-w))+ \frac{w+w^{-1}}{2}(h^n(w)-h^n(-w))\notag\\
\label{RG}
\end{eqnarray}
where $\lambda$ is a constant scale factor for each EHM step. Upon setting $w=1$, we obtain
\begin{equation}
\lambda h(1) = h(1)
\end{equation}
which implies that $\lambda=1$ unless $h(1)=0$, i.e. that the rescaling $\lambda$ for each step can be nontrivial ($\lambda \neq 1$) \emph{only if} the Hamiltonian is gapless \footnote{$k=0$ is special because this is where the UV projector has zero weight.} at the IR point $k=0$ or $w=1$. In other words, only gapless Hamiltonians can have nontrivial scale invariance, as is expected. This has very important implications in spatial dimensions $D>1$, since it implies that if the Hamiltonian contribution $h$ does not depend explicitly on $k_j$, the Hamiltonian will not be scale invariant by any EHM transformation in the $j^{th}$ dimension. 

One can derive solutions of Eq. \ref{RG} by comparing terms power by power. For instance, the second-order terms in $h(w^2)$ on the LHS force $h(w)$ to be a linear function of $w$ that is either symmetric or antisymmetric under $w\leftrightarrow w^{-1}$, i.e. a function of $w+w^{-1}$ or $w-w^{-1}$. Hence we find the two linearly independent solutions to Eq. \ref{RG} to be $h(z)= \frac{z-z^{-1}}{2i}=\sin k$ and
$h(z)=\frac{2-z-z^{-1}}{4}=\frac{1-\cos k}{2} $, both with the EHM rescaling $\lambda =\frac{1}{2}$. They are both gapless at $k=0$, as they should be, and can be combined to form the massless Dirac Hamiltonian Eq. \ref{dirac1} in $1+1$ dimensions:
\begin{equation}
H_{Dirac}(k)=v_F[\sin k \sigma_1 + M(1-\cos k)\sigma_2]
\end{equation}
where $\sigma_i$ are the Pauli matrices and $M$ controls the relative weight of the two terms. 

Note that the two terms $\sin k$ and $1-\cos k$ do not have the same scaling dimension if one takes the continuum limit $\sin k\sim k,~1-\cos k\sim k^2/2$ in ordinary RG. This illustrates the distinction of real space EHM transformation from simple momentum rescaling, due to the nontrivial influence of lattice regularization that replaces functions in $k$-space with periodic trigonometrical functions.

\section{Analytic results for $(1+1)$-dimensional boundary systems}

In Ref. \onlinecite{qi2013}, the behaviors of correlation functions and their associated bulk distances were studied numerically. In the following, we will obtain analytic results of the asymptotic behavior of bulk correlators for $(1+1)$-dim translationally-invariant boundary systems, which in turn determine the asymptotic large scale behavior of the bulk geometry we define. We will compare them with the geodesic distances between analogous points in candidate classical geometries. The higher-dimensional extensions of these results will be discussed in the next section. 

\subsection{General setup}
\label{gensetup}
We shall consider four distinct physical scenarios, all at zero chemical potential, with the representative Hamiltonian for the first three cases taken to be the Dirac Hamiltonian given in Eq. \ref{dirac1}. The results obtained should also be valid for more generic Hamiltonians, since the qualitative bulk geometry properties remain robust as long as the long distance behavior of correlators remain the same. In approximately increasing levels of sophistication, the four scenarios are:
\begin{enumerate}
\item Critical boundary Dirac model at $T=0$, corresponding to a bulk AdS (Anti-de-Sitter) geometry.
\item Massive boundary Dirac model at $T= 0$, corresponding to a ``confined geometry" with an IR termination surface.
\item Critical boundary Dirac model at $T\neq 0$, corresponding to a bulk BTZ (Ba\~nados, Teitelboim, and Zanelli) black hole geometry.
\item Critical boundary model with nonlinear dispersion at $T\neq 0$, corresponding to a bulk Lifshifz black hole geometry.
\end{enumerate}

The first two cases were already explored numerically in Ref. \onlinecite{qi2013}, with results in excellent agreement with our analytical results below.

As previously explained, the fundamental quantity to be calculated is the bulk correlator $C(n_1,n_2,\Delta (2^n x),\tau)$ given in Eq. \ref{bulkcorrelator1d}. In the thermodynamic limit $L\rightarrow \infty$, all momentum sums can be replaced by integrals:
\begin{eqnarray}
&&C(n_1,n_2,\Delta (2^nx), \tau)\notag\\
&=&\sum_q  W_{n_1}^*(q)W_{n_2}(q) e^{iq\Delta (2^nx)}G_q(\tau)\notag\\
&=&\oint_{|z|=1}\frac{dz}{z}W^*_{n_1}(z^{-1})W_{n_2}(z)z^{\Delta (2^nx)}G_{z}(\tau)\notag\\
\label{correlator1d}
\end{eqnarray}
where, as before, the conjugation symbol $^*$ indicates that only the coefficients of the polynomial $W(z)$ are complex conjugated. It is insightful to analytically continue the momentum $q$ into the \emph{complex} $z=e^{iq}$ plane, where the decay properties of the correlators can be directly read from the properties of the complex poles and branch cuts.

For comparison with the geodesic distances, we shall specialize to 2-point correlators of the following three directions in the $(2+1)$-dim bulk:
\begin{itemize}
\item Equal time, same layer ``angular'' correlator
\begin{eqnarray}
C_n(\Delta x)&=&C(n,n,\Delta x,0)\notag\\
&=&\oint_{|z|=1}\frac{dz}{z}W^*_n(z^{-1})W_n(z)z^{2^n\Delta x}G_{z}(0)
\label{correlatorx}
\end{eqnarray}
\item Equal time, different layer ``radial'' correlator
\begin{eqnarray}
C(n_1,n_2)&=&C(n_1,n_2,0,0)\notag\\
&=&\oint_{|z|=1}\frac{dz}{z}W^*_{n_1}(z^{-1})W_{n_2}(z)G_{z}(0)
\label{correlatorr}
\end{eqnarray}
\item Same site imaginary-time correlator
\begin{equation}
C_n(\tau)=C(n,n,0,\tau)=\oint_{|z|=1}\frac{dz}{z}W^*_n(z^{-1})W_n(z)G_{z}(\tau)\\
\label{correlatort}
\end{equation}
\end{itemize}
Here, we have assumed translational invariance in $x$, which is necessary for defining the correlator in terms of a Fourier integral in the angular direction.

Next we specify the boundary Hamiltonian. We shall use the Dirac Hamiltonian
\[ H_{Dirac}(k)=v_F[\sin k \sigma_1 +M(m+1-\cos k)\sigma_2]\]
from Eq. \ref{dirac1} for cases $(1)$ to $(3)$.  For the sake of conciseness in the already sundry results, we shall henceforth set $v_F$ and $M$ to unity unless otherwise stated, and consider only cases at zero chemical potential $\mu$. Indeed, $v_F$, which couples to $\tau$ under imaginary time evolution $e^{-H\tau}$ merely leads to a trivial rescaling $\tau\rightarrow v_F\tau$ in the results. The value of $M$ does not affect the leading asymptotic behavior of the correlators in general, and its study is relegated to Appendix \ref{genericdirac}. For case $(4)$, we shall simply base our calculations on the non-linear dispersion relation 
\[E_k=k^\gamma,\]
since we will be primarily interested in the effect of setting $\gamma\neq 1$. 

We next elaborate on the complex analytic structure of the Dirac Hamiltonian and correlator. The positions of the complex singularities play a crucial role in determining the \emph{asymptotic} decay behavior of the correlators, typically with power-law decay when all singularities lie on the unit circle and exponential decay otherwise. The correlator in the spatial ''angular'' direction, in particular, is a Fourier transform for which there exist results that relate the decay of Fourier coefficients with the location of singularities. For a meromorphic function $f(z)$, the Fourier coefficients $f_l = \oint_{|z|=1} \frac{dz}{z} f(z) z^l$ decay like
\begin{equation}
f_l\sim l^{-(1+B)}|z_0|^l
\label{decay}
\end{equation}
for $|z_0|<1$, $l\gg 1$, with $z_0$ the branch point of $f(z)$ closest to the unit circle and $B$ is its corresponding branching number: 
\begin{equation} f(z_0+ \Delta z) \sim f(z_0) + \left(\frac{\Delta z}{z_0}\right)^B
\label{decay2}
\end{equation}
for $z$ near $z_0$. Note that $B$ cannot be a non-negative integer, since otherwise the Riemann surface will not be ramified or even divergent at $z_0$. This result from complex analysis has been heavily used in a variety of physical problems, from the decay properties of states in condensed matter to the bulk properties of statistical networks. Proofs, together with its physical applications, can be found in Refs. \onlinecite{kohn1959,leeandy2014,leeye2015, lee2015flat} and especially \onlinecite{he2001}.

In our correlators of interest, $f(z)$ takes the explicit forms $h_z$ or $h_z/E_z$ from the expressions to follow. $h_z$ is the Dirac Hamiltonian with $M=1,v_F=1$:
\begin{equation}h_z=\left(\begin{matrix}
 & 0 & i(\frac{1}{z}-(1+m)) \\
 & -i(z-(1+m)) & 0 \\
\end{matrix}\right)\label{hz}\end{equation}
with eigenenergies
\begin{eqnarray}
E_z&=&\sqrt{1+(m+1)^2-(m+1)\left(z+\frac{1}{z}\right)}\notag\\
&  \xrightarrow[m=0]&\; \; -i(z^\frac{1}{2}-z^{-\frac{1}{2}})
\label{hz2}
\end{eqnarray}
Due to the square root, $z$ is an analytic function on a 2-sheeted Riemann surface with ramification (branch points) at $z=\infty,m+1,\frac{1}{m+1}$ and $0$. When $m=0$, the two points $z=(1+m)^{\pm 1}$ coincide and annihilate, leaving a single branch cut from $0$ to $\infty$. These branch points also appear in the flattened Hamiltonian $\frac{h_z}{E_z}$ that appears in the correlators Eqs. \ref{gq}, \ref{mueq} and \ref{gqtemp}:
\begin{eqnarray}\frac{h_z}{E_z}&=&\left(\begin{matrix}
 & 0 & \sqrt{\frac{m+1}{z}}\sqrt{\frac{z-\frac{1}{m+1}}{z-(m+1)}} \\
 & \sqrt{\frac{z}{m+1}}\sqrt{\frac{z-(m+1)}{z-\frac{1}{m+1}}} & 0 \\
\end{matrix}\right)\notag\\
&  \xrightarrow[m\rightarrow 0]&\; \left(\begin{matrix}
 & 0 & \frac{1}{\sqrt{z}} \\
 & \sqrt{z} & 0 \\
\end{matrix}\right)
\end{eqnarray}
When $m=0$ at criticality, $\frac{h_z}{E_z}$ takes a particularly simple form that does not contain any nonzero pole in the unit circle. This still holds true for much more generic critical systems, being a necessary condition for power-law decay as required by Eq. \ref{decay}. Note that all the nontrivial branch points in the integrand of the bulk correlator must come from $G(z)$, since the wavelet basis functions $W_n(z)$ or $W_n(z^{-1})$ are polynomials in $z$ or $\frac{1}{z}$.

Interestingly, this simple form of $\frac{h_z}{E_z}$ admit a real space correlator $G(\Delta x)\propto \int z^{\Delta x}\frac{h_z}{E_z}\frac{dz}{z}$ with a nontrivial phase winding. Indeed, a short calculation reveals that $G(\Delta x)\propto \frac{e^{i\pi(\Delta x \pm 1/2)/L}}{\sin\left[\frac{\pi}{L}(\Delta x + 1/2)\right]}$, which exactly agrees with numerical results for different models with a Dirac point\cite{hermanns2014}.

Although we have only explicitly studied the complex topology of the Dirac model, the important point is that more generic models, i.e. multi-band models with arbitrary dispersion also have analogous topologies that lead to similar asymptotic correlator behavior. This will be further elaborated in the last part of Appendix \ref{genericdirac}. 

Table \ref{table1} contains a summary of the asymptotic behavior of the mutual information, correlators, entanglement entropy and bulk geometry parameters for the different boundary systems. More details about each case are discussed in the remainder of this section.

\begin{table*}
\centering
\renewcommand{\arraystretch}{2}
\begin{tabular}{|l|l|l|l|}\hline
 &\ $m=T=0 $ (Case $1$) &\ $m\neq0,T=0 $ (Case $2$)  &\ $T\neq0,m=0 $ (Cases $3,4$)  \\ \hline
 $S_x$  &\ $\rightarrow \log 4$, Eq. \ref{entropy8}  &\ $\propto\frac{n}{4^{n}}\rightarrow 0$, Eq. \ref{massentropy}  &\  $\rightarrow \log 4$ \\ \hline
$I_n(\Delta x)$  &\ $\sim \frac{1}{x^6}$, Eq. \ref{Ixyangular} &\ $\sim \frac{e^{-2^{n+1}m\Delta x}2^n}{\Delta x}$,  Eq. \ref{mass2}&\ $\sim e^{-2\pi T 2^n \Delta x}$ for $\gamma=1$; and in general \\
 &\ &\  &\ $e^{-[2(\pi T)^{1/\gamma} (2^n\Delta x)]\sin \frac{\pi}{2\gamma}}$, Eq. \ref{nonlinearIxy} \\ \hline
$I(n_1,n_2)$  &\ $\sim 2^{-\Delta n}$, Eq. \ref{Ixyradial} &\ unexplored &\ unexplored \\ \hline
$\text{Tr } C_n(\tau)$  &\ $\sim \frac{2^{3n}}{v_F^3\tau^3}$, Eq. \ref{time1} &\ $\sim e^{-m\tau}$ &\ $\text{Tr }[C_n(\tau)+C_n(\beta-\tau)]_{T=0}$ for $2^n\ll 2\pi\beta$;\\ 
  &\  &\ &\ bounded below by $e^{-\frac{\beta }{2^{(2\gamma-1)n}}}$, Eqs. \ref{rhoBTZ1},\ref{rhononlinear} for $2^n> 2\pi\beta$ \\ \hline

$\frac{\xi_{\theta}}{R }$  &\ $\frac{1}{3}$ &\ n.a. &\ $1$ for $\gamma=1$ \\ \hline
$\frac{\xi_\rho}{R }$  &\ $1$ &\ n.a. &\ $1$ for $2^n\ll 2\pi \beta$; n.a. otherwise \\ \hline
$\frac{\xi_\tau}{R_\tau}$  &\ $\frac{2}{3}$ &\ n.a. &\ $\frac{2}{3}$ \\ \hline
$b$  &\ n.a. &\ n.a. &\ $L\xi_\theta T$ for $\gamma=1$; $\propto T^{1/\gamma}$ for $\gamma>1$  \\ \hline
$R_{\theta}$  &\ $\frac{1}{2}\left(\frac{1}{\pi}\sqrt{\frac{2}{I_0}}\right)^{\frac{1}{3}}$ &\ n.a. &\ not uniquely determined  \\ \hline
$R_{\tau}$  &\ $(2v_F \pi^{1/3})^{-1}$, Eq. \ref{timeR} &\ n.a. &\ $\sqrt[3]{\frac{2}{\pi^4}}$ for $\gamma=1$  \\ \hline


$\rho_n$  &\ $\frac{L}{2^{n+1}\pi}$   &\ n.a. &\ $\frac{L}{2^{n+1}\pi}$ for $2^n\ll 2\pi\beta$ \\ 
  &\  &\ &\ $b\left(1+\frac{2\beta^4}{9\pi^2 4^n}\right)$ for $2^n>2\pi\beta$  \\ \hline
\end{tabular}

\caption{Summary of results for various physical quantities in all of the $D=1$ scenarios considered, where $L$ is the size of the boundary system. Case $(1)$ ( $m=T=0$) from Sect. \ref{gensetup} is fitted onto the AdS metric with radius being either $R$ or $R_\tau$ (which are numerically close) depending on whether the fit is done for the spatial direction or imaginary time direction. The massive ($m\neq 0$) case $(2)$ is not fitted with any classical geometry. For cases $(3)$ and $(4)$ with temperature $T\neq 0$, $\gamma$ controls the dispersion via $E_q=q^\gamma$, and results are given for the BTZ case ($\gamma=1$) and the general Lifshitz case ($\gamma>1$), when applicable. The energy scale of $T$ divides the bulk into two regions demarcated by $2^n=2\pi \beta$, each with different qualitative properties. }
\label{table1}
\end{table*}

\subsection{Critical boundary Dirac model vs. bulk AdS space}

For a start, let us consider the massless $(1+1)$-dim massless Dirac model as the boundary theory. We show that the correlators in all three directions (Eqs. \ref{correlatorx} to \ref{correlatort}) all suggest a bulk geometry of a (2+1)-dim Anti de-Sitter space (further detailed in Appendix \ref{app:geodesics}) given by the metric (Eq. \ref{adsmetric}):
\begin{equation}
\frac{1+\frac{\rho^2}{R^2}}{1+\frac{L^2}{4\pi^2R^2}}d\tau^2+\frac{d\rho^2}{1+\frac{\rho^2}{R^2}}+\rho^2d\theta^2
\end{equation}
where $L$ is the boundary system size and $\tau$ is scaled such that the metric is $O(2)$-invariant at the critical boundary where $\rho=\frac{L}{2\pi}\gg R$. In the following, we shall only consider asymptotically large angular and temporal intervals with $\Delta x,\tau\rightarrow \infty$, and/or not necessarily large radial intervals between layers $1$ and $n$.  Since we have already taken the thermodynamic limit $L\rightarrow \infty$, $\Delta x<L$ can still be satisfied for arbitrarily large $\Delta x$.

\subsubsection{Spatial directions}

As is detailed in Appendix \ref{app:twoband}, the spatial decay of the Mutual Information is given by (see Eq. (\ref{Ixy2})):
\begin{equation}
I_{xy}=\frac{4(|u|^2+|v|^2)}{1-4A^2}
\end{equation}
where $|u|,|v|$ are the off-diagonal (unequal spin) part of the propagator $C_{x-y}$, and $A$ is the off-diagonal part of the onsite propagator $C_x$. From Eq. \ref{hz}, they are explicitly: 
\begin{equation}
A=-iC_n(\Delta x=0)=i\oint_{|z|=1}\frac{dz}{z}W^*_n(z^{-1})W_n(z)\sqrt{z}
\label{A}
\end{equation}
which, being onsite, does not depend on the displacement between $x$ and $y$. For the angular direction where $x,y$ are on the same layer,
\begin{equation}
u,v=-iC_n(\Delta x)=-i\oint_{|z|=1}\frac{dz}{z}W^*_n(z^{-1})W_n(z)z^{\mp \frac{1}{2}}z^{2^n\Delta x}
\label{uv}
\end{equation}
while for the radial direction where $n_1\neq n_2$ but $\Delta x=0$,
\begin{equation}
u,v=-iC(n_1,n_2)=-i\oint_{|z|=1}\frac{dz}{z}W^*_{n_1}(z^{-1})W_{n_2}(z)z^{\mp \frac{1}{2}}
\label{uv2}
\end{equation}
In the above equations, the integrands do not contain poles. However, the integrals are nonzero due to the branch cut from $z=0$ to $z=\infty$ from the square root factor. They can be evaluated by standard deformations of the contour, as demonstrated in detail in Appendix \ref{app:critical}.

After some computation, we obtain for the angular direction
\begin{equation}
|u|,|v|\sim \frac{1}{16 \pi}\frac{1}{(\Delta x)^3}
\label{uv3}
\end{equation}
As elaborated in Appendix \ref{app:critical1}, such a power-law decay of the single-particle propagator is generally expected in the presence of a branch cut. Physically, it is a signature of criticality, with a power of $3$ instead of $1$ due to the additional 'destructive interference' from the antisymmetric combinations of adjacent sites in the Haar wavelet basis. There is a striking absence of the layer index $n$ in Eq. \ref{uv3}, which reflects the scale-invariance of the boundary theory. Eq. \ref{uv3} also holds for general values of $M$ in the Dirac Model Eq. \ref{dirac1}, where $M$ controls the ratio between the quadratically dispersive $1-\cos k$ and linearly dispersive $\sin k$ terms near the IR point. While $M$ can affect the details of the branch cut, it cannot change the decay exponent, as is shown in Appendix \ref{genericdirac}.

For the radial direction with $n_1=1$ and $n_2=n$, Appendix \ref{app:critical2} also tells us that

\begin{equation}
|u|,|v|\sim \left(\frac{1}{\sqrt{2}}\right)^{n-1}
\label{uv4}
\end{equation}

Hence we have exponential decay of the single-particle propagator in the radial direction, which is consistent with scale invariance\footnote{ Scale invariance entail the multiplicative property $C(n_1,n_3)\sim C(n_1,n_2)C(n_2,n_3)$ for all $n_1<n_2<n_3$. This is only satisfied by an exponential dependence on the interval $\Delta n$.}.

Strictly speaking, the mutual information across the radial direction involves both $A_1$ and $A_n$, the unequal spins onsite propagators $A$ at layers $1$ and $n$, and a more general (and complicated) version of Eq. (\ref{Ixy2}) should be used. However, $A_n \rightarrow 0$ rapidly as $n$ increases, effectively leading to no asymptotic correction. Mathematically, this is because as $n$ increases, the peaks of $W_n(e^{iq})$ approaches a delta function at $q_0=\frac{2\pi}{2^n}\rightarrow 0$ which gets exponentially closer to the IR point, where contributions to the integral are penalized by the momentum correlator $h_q/E_q$. Explicitly, 

\begin{eqnarray}
|A_n|&=&-\frac{i}{2}\int_{-\pi}^\pi |W_n(e^{iq})|^2 \text{sgn}(q)e^{iq/2}dq\notag\\
&\rightarrow &  \frac{1}{2}\int_{-q_0}^{q_0} |W_n(e^{iq})|^2 \frac{|q|}{2}dq\notag\\
&\sim &  \frac{1}{4}|W_n(e^{iq_0})|^2\int_{-q_0}^{q_0}  |q|dq\notag\\
&=& \frac{2^{n-1}}{\pi^3}q_0^2\notag\\
&=& \frac{2}{\pi}\frac{1}{2^n}\rightarrow 0
\label{entropy1}
\end{eqnarray}
in the large $n$ limit. Physically, this means that unequal spins become totally decoupled in the IR regime. According to Eq. \ref{entropy3}, the sites in the IR layers are hence maximally entangled with the rest of the bulk:
\begin{equation}
S_x\rightarrow -4\frac{1}{2}\log \frac{1}{2}=\log{4}
\label{entropy8}
\end{equation}
This maximal entanglement in the IR also exists in generic critical systems, since the IR DOFs become harder and harder to isolate unless an energy scale (i.e. mass) exists.

Putting it all together, the mutual information behaves like
\begin{equation}
-\log I_n(\Delta x)\sim 6\log \Delta x +\log (32\pi^2)
\label{Ixyangular}
\end{equation}
for angular intervals and
\begin{equation}
-\log I(1,n)\sim (n-1)\log 2 + \text{small const.}\sim \Delta n\log 2
\label{Ixyradial}
\end{equation}
for radial intervals. These asymptotic behaviors are in excellent agreement with those of the geodesic distances on AdS space, if one uses the proposed correspondence given by $\frac{d^{min}_{\Delta \bold x}}{\xi}=- \log \frac{I_{\bold x \bold y}}{I_0}$ in Eq. \ref{dx}. The parameters $\xi$ and $I_0$ respectively set the scales of bulk distance and mutual information, and are related in a precise way discussed later. In principle, $\xi$ can be different in different independent directions.

We first study the correspondence in the angular direction. The geodesic distance between two equal-time points $(\rho,\theta_1)$ and $(\rho,\theta_2)$ with angular interval $\Delta \theta = |\theta_2-\theta_1|=\frac{\Delta x}{\rho}$ is given by Eq. \ref{ads1}:

\begin{equation}
d^{min}_{\Delta x}= R\cosh^{-1}\left(1+\frac{2\rho^2}{R^2}\sin^2\frac{\Delta \theta}{2}\right)\sim 2R \log \frac{\Delta x }{R}\label{ads1prime}
\end{equation}
where $R$ is the AdS radius that determines the length scale below which we expect significant deviations from logarithmic behavior. Comparing Eqs. \ref{Ixyangular} and \ref{ads1prime}, we see that 
\begin{equation}
\frac{R }{\xi_\theta}=3
\label{Ixy4}
\end{equation}
and
\begin{equation}
R =\frac{1}{2}\left(\frac{1}{\pi}\sqrt{\frac{2}{I_0}}\right)^{\frac{1}{3}}
\label{Ixy5}
\end{equation}
We see that $R $ decreases weakly with increasing $I_0$. For the Dirac model we take $I_0=2\log 4$, the theoretical maximal mutual information mentioned below Eq. \ref{dx}. With it, we obtain $R =0.3233$ and $\xi_\theta=0.1078$, which is in excellent agreement with the numerical values obtained in Ref. \onlinecite{qi2013}.

For the radial direction, we compare Eq. \ref{Ixyradial} with the AdS geodesic distance (Eq. \ref{ads2}):
\begin{eqnarray}
d^{min}_{\Delta \rho}&= &R |\Delta(\log \rho)|\notag\\
&=& R (\log 2) |\Delta n|
\end{eqnarray}
Here we have taken $\rho=\rho_n= \frac{L}{2\pi 2^n}$, as required by the scale-invariance of radius $\rho$ and the circumference $\frac{L}{2^n}$ at any layer $n$. We easily obtain
\begin{equation}
\xi_\rho = R 
\label{Rradial}
\end{equation}
which means that the AdS radius is nothing but the radial length scale $\xi_\rho$ of radial geodesics. This is also in agreement with the numerical results in Ref. \onlinecite{qi2013}. 

It should be noted that the ratio $R/\xi$ is different for the angular and radial directions, which is a manifestation of the fact that the mapping does not preserve the entire conformal symmetry of AdS space (since conformal symmetry is only emergent in long wavelength limit and does not exist rigorously in a lattice model). The bulk theory has scale invariance and a reduced translation symmetry with unit cell size varying with the layer index $n$, but the correlations are anisotropic between radial and angle directions.

\subsubsection{Imaginary time direction}

As previously postulated by Eq. \ref{dt}, we can deduce a classical bulk geometry from the decay properties of the imaginary time correlator $C_n(\tau)$ given by
\begin{equation}
C_{n}(\tau)=\frac{1}{2}\int_{-\pi}^\pi dq |W_n(e^{iq})|^2 e^{-\tau E_q}\left(\mathbb{I}-\frac{h_q}{E_q}\right)
\label{time0}
\end{equation}
When $\tau$ is large, the value of $C_n(\tau)$ arises from competing contributions from the IR regime and the momentum scale set by the layer index $n$. While most of the spectral weight of $W_n(e^{iq})$ is concentrated around the momentum $q_0=\frac{2\pi}{2^n}$, the exponential factor exponentially suppresses contributions above the IR  regime set by $E_q=v_F q \sim \tau^{-1}$. Hence we expect $C_n(\tau)$ to increase as $n$ goes deeper into the IR. Indeed, this is exactly contained in the analytical result Eq. \ref{time3} derived in Appendix \ref{app:time}:
\begin{equation}
\text{Tr } C_n(\tau)\sim \frac{1}{8\pi} \left(\frac{2^n}{v_F \tau}\right)^3 = \frac{1}{(8\pi^2)^2} \left(\frac{L}{v_F \rho\tau}\right)^3
\label{time1}
\end{equation}
where $\rho=\frac{L}{2\pi 2^n}$ is the bulk radial AdS coordinate of layer $n$. Eq. \ref{time1} also holds for generic $M$ in the Dirac model (Eq. \ref{dirac1}), except for the degenerate case where $M=\infty$ and there is no linearly dispersive term $\sin k\sigma_1$.

Comparing the logarithm of Eq. \ref{time1} with the AdS timelike geodesic $d_\tau \sim 2R\log \frac{2\pi\rho \tau}{LR}$ from Eq. \ref{ads3}, we obtain
\begin{equation} \frac{R}{\xi_\tau}=\frac{3}{2} \end{equation}
and
\begin{equation}
 R_\tau=\frac{1}{2v_F\pi^{1/3}}\approx \frac{0.3414}{v_F}
\label{timeR}
\end{equation}
These results are valid in the regime where $\rho\gg R_\tau$ and $R_\tau\ll \tau \ll L$, i.e. when the geodesics do not come close to circumnavigating the AdS space. The AdS radius $R_\tau$ obtained here is slightly different from $R $ obtained through spatial geodesics in Eq. \ref{Ixy5}, with the latter containing a weak dependence on the reference mutual information $I_0^{-1/6}$. 

We note that $\frac{R}{\xi_\tau}=\frac{3}{2}$ obtained from the imaginary time correlator is exactly half of that of $\frac{R}{\xi_\theta}=3$ obtained from the (spatial) mutual information. This is not an inconsistency, but a manifestation of the anisotropy in the definitions: the mutual information is asymptotically quadratic, not linear, in the two-point correlator, and its associated geodesic distance must be doubled.

\subsection{Boundary Dirac model at nonzero mass $m$, $T=0$ }

This is our first and simplest example of a non-critical system. A nonzero mass scale $m$ introduces a pole in the correlator, which leads to the exponential decay of correlation functions. As we will discuss in this subsection, the dual geometry has a spatial "termination surface" which makes the spatial geometry topologically different from that of critical fermions. The temporal geodesics, however, exhibit no unusual behavior at the termination surface, so the surface is not a black hole horizon but a purely spatial cutoff.

\subsubsection{Angular direction}

As previously discussed, the geodesics in the spatial angular direction depend on the mutual information $I_{xy}=\frac{|u|^2+|v|^2}{1-4A^2}$ where $u$ and $v$ are the unequal-spin propagators between sites separated by an angular distance of $\Delta x$ within layer $n$ , and $A$ is the unequal-spin onsite propagator. 

We first look at how $u$ and $v$ differ from those of the critical case. They are given by the off-diagonal components of
\begin{equation}
-i\oint_{|z|=1} \frac{dz}{z}W^*_n(z^{-1})W_n(z)z^{2^n\Delta x}\frac{h_z}{E_z}
\label{mass1}
\end{equation}
where $\frac{h_z}{E_z}$ is the $2\times 2$ matrix given by Eq. \ref{hz}. For the Dirac model with $m\neq 0$, $\frac{h_z}{E_z}$ and hence the integrand of Eq. \ref{mass1} now has square root branch points away from the origin, namely at $z_0=1+m$ and $z_0=\frac{1}{1+m}$. Unlike in the critical case, the singularity $z_0=\frac1{1+m}$ now satisfies $|z_0|<1$ and by Eq. \ref{decay} determines the asymptotic exponential correlator decay with $f(z)=W_n^*(z^{-1})W_n(z)\frac{h_z}{zE_z}$. The branching numbers $B$ (Eq. \ref{decay2}) at $z_0$ are given by $B=\pm \frac{1}{2}$ for $u$ and $v$, which shall thus decay (for $2^n\Delta x \gg 1$ and small $m$) like
\begin{equation}
u,v\sim \frac{(2^n\Delta x)^{-1\mp 1/2}}{(1+m)^{2^n\Delta x}}\approx \frac{e^{-m 2^n \Delta x}}{(2^n\Delta x)^{1\pm 1/2}}
\label{udecay}
\end{equation}
We see that $v$ dominates with a factor of $2^n\Delta x$, and is itself exponentially decaying with a subleading power-law decay. Hence $-\log I_{xy}$ will define a rescaled Euclidean distance with length scale given by $\frac{1}{m}$. For a Dirac model with a generic value of $M$ in Eq. \ref{dirac1}, $z_0=\frac{1}{1+m}$ should be replaced by the pole \begin{equation}z_0=\frac{M(1+m)-\sqrt{1+2mM^2+m^2M^2}}{M\pm 1},\end{equation}
whose corresponding effective mass is given by $|z_0|^{-1}-1$.

As contrasted with the critical case, the mutual information in the massive case also depends nontrivially on $A$, the unequal-spin onsite propagator, as we go deep into the IR regime below the energy scale of layer $n$. Recall that at layer $n$, $A$ is given by
\begin{equation}  A_n =  \frac{-i}{2}\int_{-\pi}^{\pi}|W_n(e^{iq})|^2 \frac{h_q^{offdiag}}{E_q} dq \label{entropy5}
\end{equation}
For large $n$, the spectral weight is mainly concentrated around $q_0=\frac{2\pi}{2^n}$, where
\begin{eqnarray} \frac{h^{offdiag}_q}{E_q} &=& \frac{\sin q \pm i(m+1-\cos q)} {E_q}\notag\\
&\rightarrow& \frac{i(m+\frac{q^2}{2})}{\sqrt{m^2+q^2}}\notag\\
& \approx& i\left(1+\frac{m-1}{2m^2}q^2 \right)
\label{hE}
\end{eqnarray}
where the $\sin q$ term had been dropped because it is odd. Hence Eq. \ref{entropy5} can be rewritten as
\begin{eqnarray}
|A_n|&=&\frac{1}{2}\int_{-\pi}^\pi |W_n(e^{iq})|^2 \left(1+\frac{m-1}{2m^2}q^2 \right)dq\notag\\
&\rightarrow &  \frac{1}{2}+\frac{m-1}{4m^2}\int_{-q_0}^{q_0} |W_n(e^{iq})|^2q^2 dq\notag\\
&\sim &  \frac{1}{2}-\frac{1}{4m^2}|W_n(e^{iq_0})|^2\int_{-q_0}^{q_0}  q^2 dq\notag\\
&=& \frac{1}{2}-\frac{2^{n}}{8m^2\pi^3}\frac{q_0^3}{3}\notag\\
&=& \frac{1}{2}-\frac{1}{3m^2}\frac{1}{4^n}
\label{massa}
\end{eqnarray}
This leads to an overall enhancement of $\frac{1}{1-4A_n^2}\sim 4^{n}$ in the mutual information, which thus behaves like
\begin{equation}
-\log I_n(\Delta x) \sim 2^{n+1}m\Delta x + \log \Delta x - n\log 2
\label{mass2}
\end{equation}
Its leading contribution is proportional to $2^n\Delta x$, which leads to a  key difference of its dual geometry from that of the critical fermion. Since the corresponding geodesic distance $d_{\Delta x}=\xi_{\Delta x}I_n(\Delta x)$ is linear in $\Delta x$, i.e. like an Euclidean distance, we can obtain the circumference $\alpha_n$ of layer $n$ simply by taking $\Delta x=2^{N-n}$, the number of sites in the whole layer. More rigorously, the circumference $\alpha_n$ should be measured by first taking equally spaced points with $2^{N'}$ sites between them, and then taking the $N\rightarrow \infty$ limit before taking the $N'\rightarrow \infty$ limit. The circumference will be given by the product of the geodesic distance between neighboring points $2m\xi_{\Delta x}\cdot 2^{n+N'}$ and the number of points $2^{N-n-N'}$, i.e. $\alpha_n=2m\xi_{\Delta x}\cdot 2^N$ which is a finite portion of the boundary circumference $2^N$.

\begin{figure}[H]
\includegraphics[scale=.42]{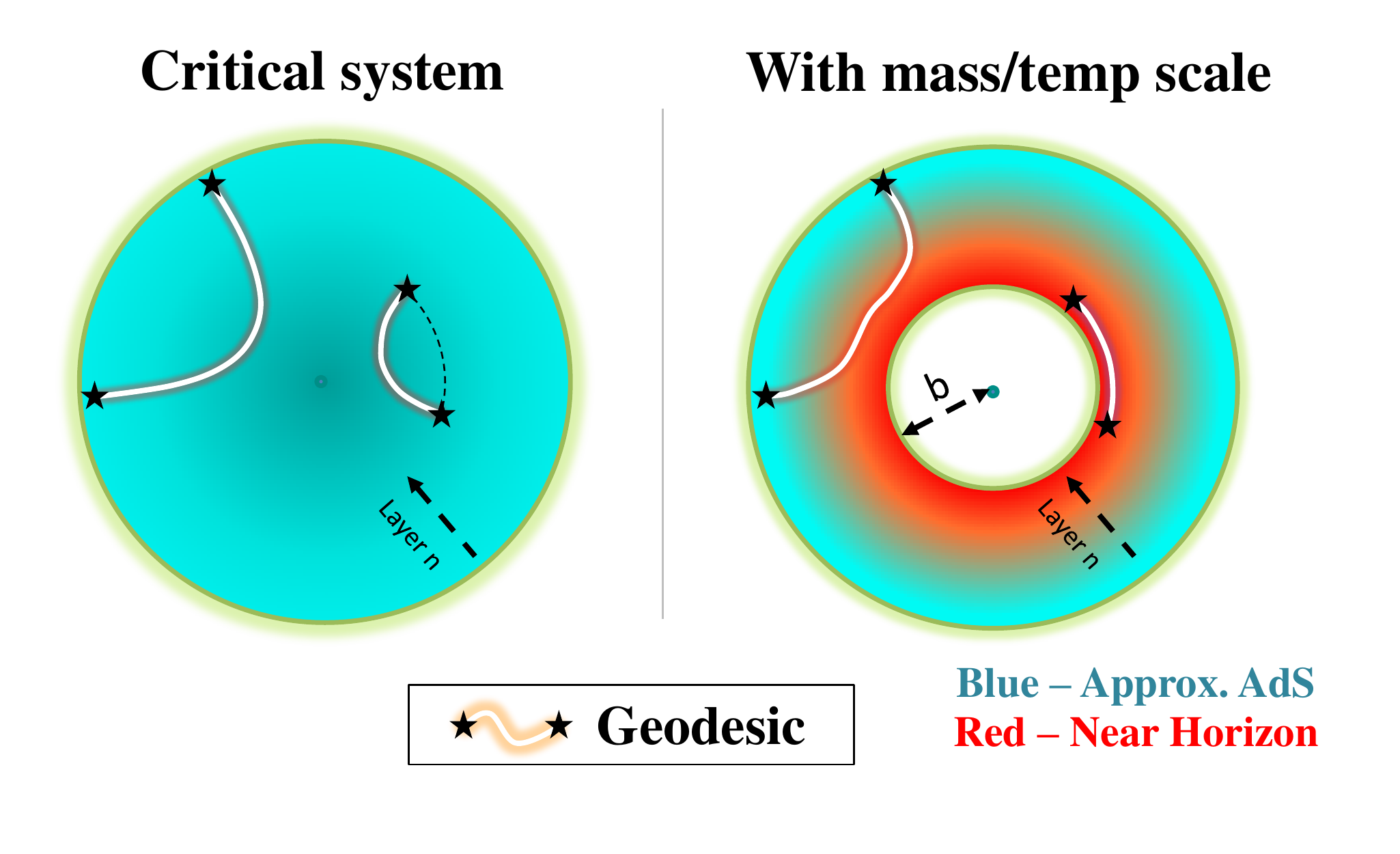}
\caption{(Color Online) The spatial bulk geometries of critical (Left) and non-critical (Right) systems and examples of their geodesics. In the AdS bulk (Left) corresponding to a critical boundary system, geodesics extend inwards towards the center, where there is far less 'space'. With an energy scale introduced by nonzero mass or temperature, the bulk geodesics develops a spatial 'termination' of space (Right). Geodesics wrap around a 'prohibited' region of radius $\propto m$ or $T$, acquiring an Euclidean character is that regime. As explained further in the main text, only the case at nonzero temperature corresponds to a true horizon. }
\label{fig:ADS_BTZ}
\end{figure}

The fact that $\alpha_n$ is finite in the large $n$ (infrared or IR) limit tells us that when $n\rightarrow \infty$, we are not approaching the center of a hyperbolic disk, but rather approaching a surface with finite area which acts as a ``termination surface" of the space. This is illustrated in Fig. \ref{fig:ADS_BTZ}. The IR surface shrinks with decreasing mass $m$. By comparison, in the dual geometry corresponding to the critical fermion, the distance between points $\Delta x$ sites apart does not depend on the layer index, so that the circumference $\alpha_n\propto 2^{-n}$ decays exponentially in the IR limit.

It is important to note that this surface is not a black-hole horizon, which distinguishes it from the case of nonzero  temperature state we will discuss in subsequent sections. One evidence for this conclusion is that each site carries a vanishing entanglement entropy $S_x$ in the IR limit $n\rightarrow \infty$, since
\begin{eqnarray}
S_x
&\approx& -\frac{2}{3m^2}\frac{1}{4^n}\left(\frac{1}{3m^2}\frac{1}{4^n}+\log\left(\frac{1}{3m^2}\frac{1}{4^n}\right)-1\right)\notag\\
&\sim& \frac{ \log 16}{3m^2}\frac{n}{4^n}
\label{massentropy}
\end{eqnarray}
which is obtained by substituting Eq. \ref{massa} into Eq. \ref{entropy3}. Since the entropy decays exponentially at large $n$, the IR sites actually for direct product states unentangled with one  another. This is consistent with the physical picture that mass is renormalized to exponentially larger values in the IR limit (with respect to the kinetic energy scale), which forces the IR ground state to be simply a direct product of the single-site ground states with a large mass term. 

\subsubsection{Imaginary time direction}

The correlator in the imaginary time direction can be obtained pretty straightforwardly. Substituting the expression for the energy dispersion $E_q=\sqrt{\sin^2q+(1+m-\cos q)^2}\approx m+\left(\frac{1+m}{2m^2}\right)q^2$ into Eq. \ref{time0}, we obtain
\begin{eqnarray}
\text{Tr } C_{n}(\tau)&=&e^{-m\tau}\int_{-\pi}^\pi dq |W_n(e^{iq})|^2 e^{-\tau \frac{1+m}{2m^2}q^2}\notag\\
&\propto& e^{-m\tau}
\end{eqnarray}
which is an exponentially decaying term multiplied by a nonuniversal Gaussian Integral. Hence $-\log \text{Tr } C_{n}(\tau)\sim m \tau$ defines an Euclidean bulk geodesic in the imaginary time direction. Together with the  previous results on spatial geometry, we see that although there is a spatial termination surface in the IR limit, the temporal direction still extends as usual. This is consistent with our earlier statement that the IR termination surface is not a black hole horizon, since the time direction is not infinitely redshifted. Topologically, the space-time we obtain for the massive Dirac model is $\mathcal{R}\times \mathcal{M}$, where $\mathcal{R}$ is the line in time direction, and $\mathcal{M}$ is a spatial annulus. 

\subsection{Critical linear boundary Dirac model at nonzero temperature $T$ vs. bulk BTZ black hole}
\label{sec:temp}

Now we study the effect of nonzero temperature in the massless Dirac model, the bulk geometry dual to which will be interpreted as a BTZ (Ba$\tilde{n}$ados, Teitelboim and Zanelli) black hole\cite{banados1992}. The BTZ black hole is a black hole solution for $(2+1)$-dim gravity with a negative cosmological constant, whose metric is
\begin{equation}
ds^2=\frac{V(\rho)}{V(L/(2\pi))}d\tau^2 + \frac{d\rho^2}{V(\rho)} +\rho^2 d\theta^2
\label{BTZmetric}
\end{equation}
where $V(\rho)=\frac{\rho^2-b^2}{R^2}$ is the Lapse function, $b$ is the horizon radius and $R$ is an overall length scale. We have rescaled $\tau$ by a factor of $\left[V\left(\frac{L}{2\pi}\right)\right]^{-1/2}$ so that $ds^2\rightarrow d\tau^2+\rho^2d\theta^2$ possesses $O(2)$ symmmetry at the boundary where $2\pi\rho =L$. The rescaling makes $\tau$ the same imaginary time variable as that in the boundary system (the canonical conjugate coordinate to the boundary Hamiltonian). 

Most notably, we find that the layers deep in the IR accumulate at the bulk horizon radius $\rho\rightarrow b$ from above, where $b$ is proportional to the temperature $T=\beta^{-1}$. This result is rigorously accurate in the $m\sim O(1)$ range of the Dirac model given by Eq. \ref{dirac1}, where the critical dispersion is essentially linear. 

\subsubsection{Angular direction}

Like a nonzero mass, a nonzero temperature also introduces an energy scale into the system. This energy scale is manifested as an imaginary gap $-log |z_0|>0$ where $z_0$ is the singularity of the momentum-space correlator $z_0$ closest to the unit circle. In this case, the singularities originate from the $\tanh\frac{\beta E_q}{2}$ term\footnote{The $\tanh\frac{\beta E_q}{2}$ term also cancels the singularity from $\frac{1}{E_q}$ on the unit circle, hence opening up the possibility of faster-than-power-law decay.} given in Eq. \ref{gqtemp}.

Above the energy scale of $T$, all correlators and hence the mutual information do not feel the effect of thermal excitations, and define an approximate AdS geometry just as in the $T=0$ critical case.

As one goes below the energy scale of $T$, Eq. \ref{decay} states that the unequal-spin propagators $u,v$ decay exponentially as $|z_0|^{2^n\Delta x}$, with a subleading power-law term $(2^n\Delta x)^{-(1+B)}$. $B$ and $|z_0|$ can be found as follows. 

Expanding $\cosh\frac{\beta E_z}{2}$, the denominator of the singular term, about $z_0$, we find that
\begin{equation}
 \cosh\frac{\beta E_z}{2}=0+\frac{\beta(z-z_0)}{2}\sinh\frac{\beta E_z}{2} \frac{dE_z}{dz} \propto z-z_0
\label{powerdecay}
\end{equation}
so the branching number $B=-1$. This value of branching number holds universally for generic systems at nonzero temperature\footnote{Except in the rare case where the energy scales associated with the mass and temperature exactly coincide. Singularities due to massive branch points diverge at zeros of $E_z$, unlike in the case of nonzero  temperature, and may consequently possess a fractional $B$. }, since Eq. \ref{powerdecay} does not depend on the form of $E_z$. Therefore, there is \emph{rigorously no} subleading power-law term in the decay of correlators in the nonzero  temperature case.

We now proceed to find $|z_0|$. $\tanh\frac{\beta E_q}{2}$ is singular when $\cosh \frac{\beta E_q}{2}=0$, i.e. $i\pi=\beta E_q =-i\beta(z^{1/2}-z^{-1/2})$ (see Eq. \ref{hz2}). This is satisfied by
$z+\frac{1}{z}-2=\left( \frac{2\pi(2l+1)}{\beta}\right)^2$, where $l\in \mathbb{Z}$. Hence the poles occur at
\begin{eqnarray}
 z_0^\pm&=&1+\frac{\pi^2 (2l+1)^2}{2\beta^2}\pm \frac{\pi(1+2l)\sqrt{4\beta^2+(2l+1)^2\pi^2}}{2\beta^2}\notag\\
&\rightarrow& 1\pm \pi\frac{2l+1}{\beta}
\end{eqnarray}
in the limit of small $T=\frac{1}{\beta}$. The pole with $l=0$ is closest to the unit circle, and we thus have the asymptotic decay
\begin{equation}
u\sim v\sim |z_0^-(l=0)|^{2^n\Delta x}\rightarrow(1-\pi T)^{2^n\Delta x}\approx e^{-\pi T 2^n \Delta x}
\label{udecay2}
\end{equation}
so that
\begin{equation}
I_n(\Delta x)\sim \frac{8e^{-\pi T 2^{n+1} \Delta x}}{1-4A_n^2}
\label{udecay3}
\end{equation}
Since $A_n\rightarrow \frac{1}{2}$ after the first few $n$, as explained in the subsection on the zero temperature critical case $(1)$, the mutual information behaves like (recalling that $2^n\Delta x = \frac{L\Delta \theta}{2\pi}$)
\begin{equation}
-\log I_n(\Delta x)\sim 	2\pi T 2^n\Delta x =LT \Delta \theta
\label{temp1}
\end{equation}
with no logarithmic subleading term. This asymptotic form for $I_n\left(\Delta x\right)$ is, to leading order, the same as that of Eq. (\ref{mass2}) for the massive zero temperature case, if we replace $2m$ by $2\pi T$. Following along the same lines as the previous section, we conclude that the circumference of each circle at layer $n$ is asymptotically $\alpha_n\simeq 2\pi T\cdot 2^N$. Thus there is a termination surface with this circumference (1-dim area) in the IR (low energy) limit. However, as we will verify by the single-site entropy and imaginary time direction distance later in this section, this surface is not just a termination surface of space, but a black hole horizon. Before discussing that, we shall first make a detailed comparison of the angular direction distance defined by Eq. \ref{temp1} with the angular geodesic distance given by a BTZ black hole metric. 

We can obtain a precise relationship between the temperature $T$ and the black hole radius $b$. Since Eq. \ref{temp1} holds in the IR regime, we compare it with the BTZ geodesic distance\footnote{Intuitively, the geodesic distance is strictly linear in $\Delta x$ infinitesimally near the horizon because the geodesics are not allowed to have any radial extent. This occurs when $V(\rho)\rightarrow 0$, which results in a radial displacement $\Delta s= \frac{\Delta \rho}{V(\rho)}$ becoming infinitely costly. } $d^{min}_{\Delta x}=2R \sinh^{-1}\left[\frac{\rho}{b}\sinh\frac{b\Delta \theta}{2R}\right]$ from Eq. 23 of Ref. \onlinecite{qi2013} in the near horizon limit $0<\rho-b\ll b$. 
\begin{eqnarray}
-\log I_n(\Delta x)= \frac{d}{\xi_\theta}&=& \frac{2R}{\xi_\theta} \sinh^{-1}\left[\frac{\rho}{b}\sinh\frac{b\Delta \theta}{2R}\right]\notag\\
&\approx& \frac{2R}{\xi_\theta} \log\left(\frac{\rho}{b}e^{b\Delta \theta/2R}\right)\notag\\
&=&\frac{ 2R}{\xi_\theta} \log\frac{\rho}{b}+\frac{2^{n+1}b \pi \Delta x}{\xi L}\notag\\
&\approx &\frac{2 \pi b(2^n \Delta x)}{\xi_\theta L}
\label{BTZcompare1}
\end{eqnarray}
which implies that
\begin{equation} b=L\xi_\theta T
\label{T1}
\end{equation}
Here $\xi_\theta$ is a yet-undertermined length scale. Our bulk geometry has agreed remarkably well with that of an actual BTZ black hole, with the LHS and RHS of Eq. \ref{BTZcompare1} agreeing on not just the leading term linear in $\Delta x$, but also the \emph{vanishing} logarithmic subleading term. Further discussions of the near-horizon geometry can be found in Appendix \ref{app:rindler}.

$\xi_\theta$ can be determined by compairing Eq. (\ref{T1}) with the relation between $T$ and $b$ in classical gravity. The requirement that the geometry is smooth in imaginary time at $\rho=b$, i.e. without a conical singularity, requires\cite{gibbons1977,charmousis}
\begin{eqnarray}
T&=& \left(\frac{2\pi R^2}{b}\frac{L}{2\pi R}\right)^{-1}= \frac{b}{RL}
\label{T2}
\end{eqnarray}
More details about this formula are given in Appendix \ref{app:rindler}. Comparing Eqs. (\ref{T1}) and (\ref{T2}) we obtain
\begin{equation} R=\xi_\theta
 \end{equation}
so that $\xi_\theta$ and $R$ are actually one and the same length scale parameter.

\subsubsection{Imaginary time direction}

The nonzero temperature features prominently in the imaginary time correlator because a finite $\beta$ corresponds to a finite periodicity in imaginary time. We shall show that the correlator $\text{Tr } C_n(\tau)$ defines a bulk geometry that is qualitatively similar to that of BTZ black hole, and deduce the effective radius $\rho=\rho_n$ of layer $n$ by looking at the maximal value of $-\log \text{Tr } C_n(\tau)$ achieved at half-period $\tau=\frac{\beta}{2}$.

The quantity to be computed is
\begin{equation}
\text{Tr } C_n(\tau)= \frac{1}{2}\int^{\pi}_{-\pi} dq |W_n(e^{iq})|^2 \text{Tr } G_q(\tau)
\end{equation}
where, from Eq. \ref{gq0},
\begin{equation}
\text{Tr } G_q(\tau) = \sum_{\lambda_q}\frac{e^{\lambda_q \tau}}{1+e^{\beta \lambda_q}}=\frac{1}{2}\sum_{\lambda_q}\frac{e^{\eta \lambda_q }}{\cosh \frac{\beta \lambda_q}{2}}=\frac{\cosh\eta E_q }{\cosh \frac{\beta E_q}{2}}.
\label{gqD}
\end{equation}
where $\eta=\tau-\frac{\beta}{2}$ and $\lambda_q$ denotes an eigenenergy of the system. The intermediate steps of Eq. \ref{gqD} are valid for any number of bands, but we have specialized to $\lambda_q=\pm E_q$ for the particle-hole symmetric 2-band case at the last step. $\text{Tr }G_q(\tau)$ is manifestly even in $\eta$, which guarantees that $\text{Tr } G_q(0)=\text{Tr } G_q(\beta)$. It can also be obtained from Eq. \ref{gq} through direct simplification.

Due to the presence of the energy scale set by $T$, there are two distinct regimes. In the (high energy) UV limit $2^n\ll 2\pi \beta$ the kinetic energy dominates the temperature, and we have $\frac{\cosh\eta E_q }{\cosh \frac{\beta E_q}{2}}\approx e^{-\tau E_q}+e^{(\tau-\beta)E_q}$, yielding the nice relation
\begin{equation}
\text{Tr } C_n(\tau)\approx \text{Tr } C_n(\tau)|_{T=0}+\text{Tr } C_n(\beta-\tau)|_{T=0}.
\label{timetemp}
\end{equation}
This equation tells us that the nonzero  temperature correlator above the energy scale of $T$ is just the superposition of two copies of the zero temperature correlator reflected about $\tau=\frac{\beta}{2}$. This is consistent with how correlation functions in BTZ geometry are obtained by a periodic quotient of those in AdS space\cite{lifschytz1994}. Eq. \ref{timetemp} is completely general, since we haven't used any particular form for the Hamiltonian.

\begin{figure}[H]
\includegraphics[scale=.8]{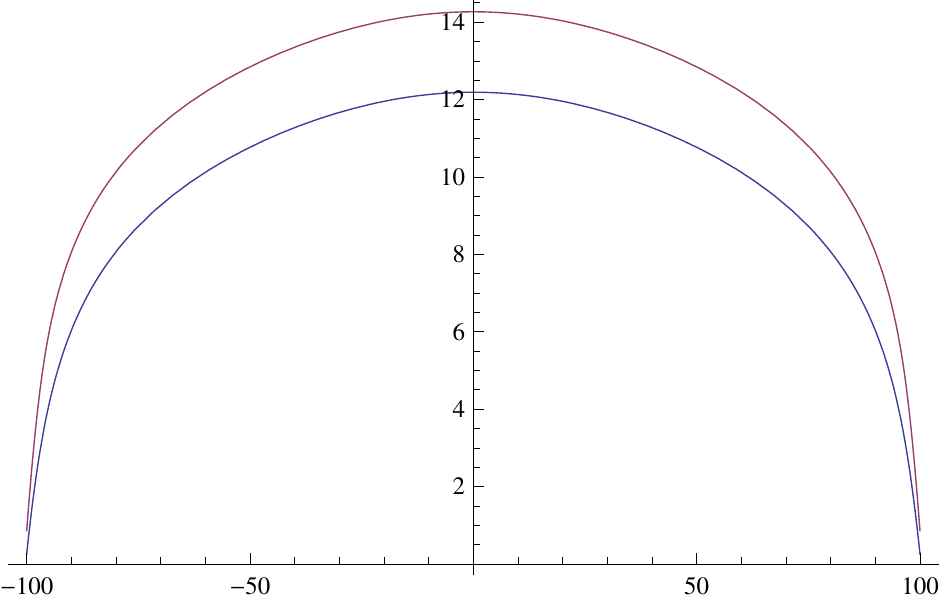}
\caption{$-\log\text{Tr }C_n(\tau)$ according to Eq. \ref{timetempD}. Plotted are the $n=1,2$ curves for  $T=0.005$, which corresponds to an energy scale of $n\approx 10$. }
\label{periodic_time}
\end{figure}

Inserting the result of $\text{Tr } C_n(\tau)|_{T=0}$ from Eq. \ref{time1}, we find that
\begin{eqnarray}
&&-log  \;\text{Tr } C_n(\tau)\notag\\
&=& -\log\left(\tau^{-3}+(\beta-\tau)^{-3}\right)-3(n-1)\log2 +\log (\pi )\notag\\
\label{timetempD}
\end{eqnarray}
As suggested in Fig. \ref{periodic_time}, the first term gives a curve with geodesic distance qualitatively similar to that outside a BTZ black hole where $\rho\gg b$:
\begin{eqnarray}
d_\tau&=& R\cosh^{-1}\left[\frac{2\rho^2}{b^2}\sin^2\frac{\pi\tau}{\beta}\right]
\label{dtau}
\end{eqnarray}
which is derived in the Appendix of Ref. \onlinecite{qi2013}.

We can find the length scales $R$ and $\xi_\tau$ by comparing the maximal value of $-\log \text{Tr } C_n(\tau)$ with $\frac{d_\tau}{\xi_\tau}$, $\tau=\frac{\beta}{2}$. We have
\begin{eqnarray}
\frac{d_{\beta/2}}{\xi_\tau}&=& \frac{R}{\xi_\tau}\cosh^{-1}\frac{2\rho^2}{b^2}\notag\\
&\approx & \frac{2R}{\xi_\tau}\log\frac{2\rho}{b}\notag\\
&=& \frac{2R}{\xi_\tau}\log\frac{2\rho\beta}{LR}\notag\\
&\approx& \frac{2R}{\xi_\tau}\log\frac{\beta}{R\pi 2^n}
\end{eqnarray}
where we have used $\rho=\rho_n\approx \frac{L}{2\pi2^n}$ in the approximate AdS geometry far from the horizon. Comparing this with $-\log \text{Tr } C_n(\beta/2)= 3\log\frac{\beta }{2^n}+\log\frac{\pi}{2}$, we obtain
\begin{equation}
\frac{R}{\xi_\tau}= \frac{3}{2}
\end{equation}
which is exactly the same as in the $T=0$ case. This must be true because the bulk geometry is still asymptotically AdS above the energy of $T$. However, the value of $R$ is somewhat different due to the different functional forms of Eqs. \ref{timetempD} and \ref{dtau}, and takes the value
\begin{equation}
R=\sqrt[3]{\frac{2}{\pi^4}}\approx 0.274
\end{equation}
In the IR limit below the energy scale of $T$, i.e. $2^n> 2\pi\beta$, Eqs. \ref{timetemp} and hence \ref{timetempD} no longer hold because we must use the small $\beta E_q$ approximation in the denominator of $\text{Tr }G_q(\tau)=\frac{\cosh{\eta E_q}}{\cosh \frac{\beta E_q}{2}} $. At $\tau=\frac{\beta}{2}$ or $\eta=0$, we have, to 2nd order in $\beta E_q \ll 1$,
\begin{eqnarray}
\text{Tr } G_q\;(\tau=\beta/2)&\approx & \sum_{\lambda_q}\left(1-\frac{\beta^2 \lambda_q^2}{8}\right)\approx  1-\frac{\beta^2}{8}q^2
\label{rhoBTZ0}
\end{eqnarray}
Hence,
\begin{eqnarray}
&&-\log \text{Tr } C(\beta/2)\notag\\
&=& -\log \text{Tr }\int dq  G_q(\beta/2) |W_n(e^{iq})|^2\notag\\
&=& -\log\left(\int dq  |W_n(e^{iq})|^2 - \frac{\beta^2 }{8}\int dq q^2 |W_n(e^{iq})|^2\right)\notag\\
&\approx & -\log\left(1 - \frac{\beta^2 }{8}q_0^2  |W_n(e^{iq_0})|^2\right)\notag\\
&\approx & \frac{\beta^2 }{2^n\pi}
\label{rhoBTZ1}
\end{eqnarray}
where we have used the fact that $W_n(e^{iq})$ is sharply peaked at $q_0=\frac{2\pi}{2^n}$ with magnitude $\sqrt{\frac{2^{n+1}}{\pi^3}}$. Comparing Eq. \ref{rhoBTZ1} to the BTZ geodesic distance in Eq. \ref{dtau}:
\begin{equation} -\log \text{Tr } C(\beta/2)\sim \frac{2R_\tau}{\xi_\tau}\cosh^{-1}\frac{\rho}{b}\approx \frac{3}{2}\sqrt{2\left(\frac{\rho}{b}-1\right)}
\end{equation}
we arrive at
\begin{equation}
\rho_n\propto b\left(1+\frac{2}{9\pi^2}\frac{\beta^4 }{2^{2n}}\right)
\label{rhoBTZ}
\end{equation}
Indeed, $\rho\rightarrow b$ exponentially from above. This justifies the previous assumption that $\frac{\rho}{b}-1\ll 1$ which is, among other implications, consistent with the fact that no non-negligible logarithmic subleading terms exist in Eq. \ref{BTZcompare1}. Physically, the agreement between the imaginary time distance of the bulk geometry and the BTZ black hole means that the rate of imaginary time correlator decay slows down exponentially as one goes deeper into the IR, which translates into the infinite redshift an outside observer sees for any physical process near the BTZ horizon. 

\subsection{Critical boundary model with nonlinear dispersion vs. bulk Lifshitz black hole}

As a sequel to the previous subsection, we now consider a nonzero  temperature boundary system with nonlinear dispersion in the long wavelength limit. As we have seen in the previous subsection, the energy-momentum dispersion is sufficient for determining the decay properties of the correlators. Here we consider the simplest nonlinear critical dispersion
\begin{equation}
E_q\simeq q^\gamma,\text{~for~}q\approx  0
\label{eqgamma}
\end{equation}
For higher $q$, $E_q$ should be regularized to a periodic function. However, the details of the regularization do not affect the critical behavior as we have seen time and again, and Eq. \ref{eqgamma} is sufficient as it stands. In this sense, Eq. \ref{eqgamma} subsumes the Dirac model (Eq. \ref{dirac1}) with general $M$, which is merely an interpolation between a $\gamma=1$ and a $\gamma=2$ Hamiltonian.

The asymptotic behavior of the mutual information depends solely on the position of the singularities of $G_q(\tau)\propto \text{sech}\frac{\beta E_q}{2}$. They occur when $\beta E_q=i(2l+1)\pi$, $l\in\mathbb{Z}$, i.e. when
\begin{equation}
E_q^2=q^{2\gamma}=-\pi^2 T^2(2l+1)^2
\end{equation}
In terms of $z=e^{iq}$, they occur at $z_0=e^{i\left(i\pi T (2l+1)\right)^{1/\gamma}}$. Clearly, the $l=0$ singularity has the largest magnitude within the unit circle, which is given by
\begin{equation}
|z_0|=e^{-(\pi T)^{1/\gamma}\sin \frac{\pi}{2\gamma}}
\end{equation}
From Eq. \ref{decay} and discussions surrounding Eqs. \ref{udecay3}, we conclude that the mutual information between two points with angular separation of $\Delta x$ sites decay like
\begin{eqnarray}
-\log I_n(\Delta x) &\sim& 2(\pi T)^{1/\gamma}(2^n\Delta x)\sin \frac{\pi}{2\gamma} \notag\\
&=&\frac{L\Delta \theta}{\pi}(\pi T)^{1/\gamma}\sin \frac{\pi}{2\gamma}
\label{nonlinearIxy}
\end{eqnarray}
Similar to case $(3)$ with linear dispersion, the $\Delta \theta$ dependence in the mutual information suggests that the circumference approaches a finite value in the IR limit, so that there is an event horizon. However, the nonlinear dispersion leads to a different $T$ dependence. Since the bulk geodesic distance $\propto - \log I_n(\Delta x)\propto T^{1/\gamma}\Delta \theta $, the black hole radius is $b\propto T^{1/\gamma}$, different from the $b\propto T$ behavior of the BTZ black hole\footnote{This is incompatible with metrics with Galilean symmetry, i.e the BTZ metric in Eq. \ref{BTZmetric}, because they always require $b\propto T$ to avoid a conical singularity in Rindler space.}.

There are indeed black hole solutions to classical gravity with the property $b\propto T^{1/\gamma}$, if one considers spacetimes with anisotropic scale invariance, i.e. with metric\cite{ayon2009}
\begin{equation}
ds^2=-\frac{r^{2\gamma}}{R^{2\gamma}}dt^2 + \frac{R^2}{r^2}dr^2 + \frac{r^2}{R^2}d\vec x^2
\end{equation}
invariant under the rescaling $(t,\vec x,r)\rightarrow (\lambda^\gamma t, \lambda \vec x, \lambda^{-1}r)$, with $\gamma$ the dynamical critical exponent and $R$ a length scale. 

If we include quadratic curvature tensor terms like $\Omega^2$, $\Omega_{\alpha\beta}\Omega^{\alpha\beta}$ or $\Omega_{\alpha\beta\mu\nu}\Omega^{\alpha\beta\mu\nu}$ to the gravitational action\cite{ayon2009,ayon2010}, the resultant Einstein's equations in the non Galilean-invariant spacetime will possess black hole solutions for certain ranges of parameters. Such solutions are known as \emph{Lifshitz Black Holes}, which are well-studied\cite{ayon2009,ayon2010, balasubramanian2009, cai2009} and proposed as possible gravity duals to Lifshitz fixed points in condensed matter physics. The explicit solutions of these black holes are known for certain values of $\gamma$, especially in the $2+1$ dimensions relevant to our current context.  For instance, the black hole metric for $\gamma=3$ and the gravitational action $S=\frac1{16\pi G}\int d^3x \sqrt{-g}\left[\Omega -2 \Gamma +2R^2\left(\Omega_{\alpha\beta}\Omega^{\alpha\beta}-\frac{3}{8}\Omega^2\right)\right]$, where $\Gamma$ is the cosmological constant, is given by\cite{ayon2009}
\begin{equation}
ds^2=-\frac{\rho^6}{R^6}\left(1-\frac{b^2}{\rho^2}\right)dt^2+\frac{R^2d\rho^2}{\rho^2-b^2}+r^2d\theta^2
\label{lifshitzmetric}
\end{equation}
By examining the near-horizon geometry in Euclidean time, we explicitly find its Hawking temperature to be $T=\frac{b^3}{2\pi R^4}$, which agree with the horizon area we obtained from a boundary theory with cubic dispersion.

The $T\propto b^\gamma$ dependence can also be expected from a simple counting argument. A system at a temperature $T$ can be physically understood as one with states randomly distributed in a energy width of $T$. This randomness is quantified by the entanglement entropy $S$ of the system with the thermal bath, most of which is carried by the IR region of the system. In the bulk system obtained through the EHM, the IR states carry maximal entropy per each site, so that the thermal entropy is proportional to the number of sites in the ``stretched horizon", which is proportional to the horizon area $b$. For a system with energy dispersion $E\propto q^\gamma$, the momentum range that has energy below $T$ is $\Delta k\propto T^{1/\gamma}$, so that the entropy $S\propto \Delta k\propto T^{1/\gamma}$. Consequently, $b\propto T^{1/\gamma}$. 

The imaginary time bulk correlator behaves in a similar way as that with linear dispersion (Case $(3)$). For the layers with energy scale above $T$, we still have, of course,
\[\text{Tr } C_n(\tau)=\text{Tr }[C_n(\tau)+C_n(\beta-\tau)]_{T=0}\]
which holds independently of the dispersion. In the IR limit with energy scale below $T$, the minimal value for $\text{Tr } G_q(\tau)$ at $\tau=\frac{\beta}{2}$ still follows from Eq. \ref{rhoBTZ0} and \ref{rhoBTZ1}, except that the energy eigenvalues are now $\lambda=q^\gamma$. Hence we obtain a nontrivial (but still simple) nonlinear correction
\begin{equation}
-\log \text{Tr } C(\beta/2)\sim \frac{\beta^2}{2^{(2\gamma-1)n}\pi}
\label{rhononlinear}
\end{equation}
which is consistent with the metric in Eq. \ref{lifshitzmetric}, which has the geodesic distance at $\tau=\frac{\beta}{2}$ vanishing as a power of $\frac{\rho-b}{b}$. We still have $\rho\rightarrow b$ exponentially as $n$ increases, but at a different rate compared to the BTZ (Galilean-invariant) case.

\section{Generalization of EHM to higher dimensions}
\label{sec:higherdim}

\subsection{General setup}

\begin{figure}[H]
\centering
\includegraphics[scale=.4]{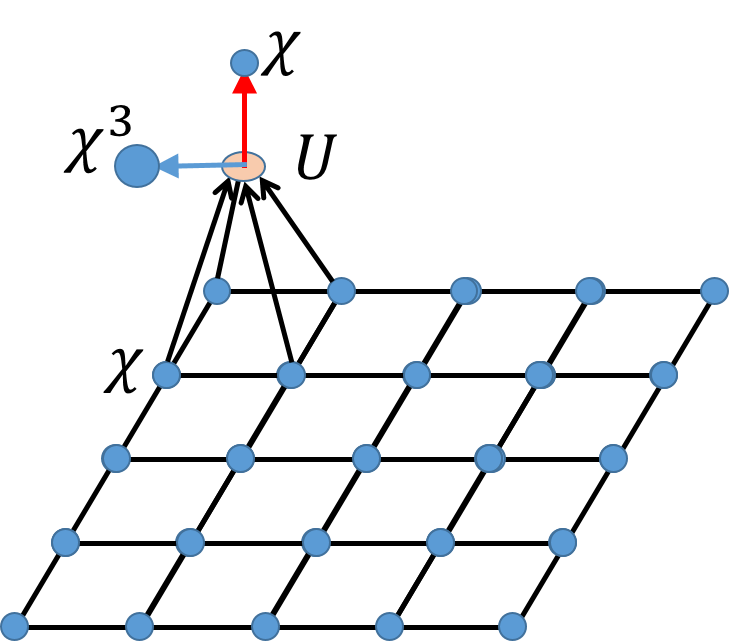}
\caption{Schematic picture of the EHM for two-dimensional boundary states. A unitary mapping (pink oval) is defined for four sites, each with a $\chi$-dimensional Hilbert space. It maps the DOFs from the four sites to two output sites, the IR site (red arrow) with dimension $\chi$ and the UV site (blue arrow) with dimension $\chi^3$. }
\label{fig:EHM2d} \end{figure}

When we generalize the boundary system to $D$ spatial dimensions, the bulk system will contain $D+1$ spatial dimensions, with a new emergent direction representing the energy scale. The boundary theory and bulk theory can be related by EHM in the same way as in the $D=1$ case. For example, with a boundary theory defined on the two-dimensional square lattice, a unitary mapping can be defined on four sites around a plaquette, which maps it to two sites representing the high energy and low energy degrees of freedom of the four sites. The mapping is illustrated in Fig. \ref{fig:EHM2d}. If the Hilbert space dimension is $\chi$ on each site, the output IR site should have dimension $\chi$ while the UV site now has a higher dimension $\chi^3$, corresponding to the $UV,IR$, $IR,UV$ and $UV,UV$ sectors of the 1-dim case. More generally in $D$ dimensions with a square lattice, one can map the $2^D$ sites in a cube to one IR site with Hilbert space dimension $\chi$ and one UV site with dimension $\chi^{2^D-1}$.

For free fermions systems, the mapping is equivalent to a wavelet transformation on the single-particle wavefunctions, just like in the case with one spatial dimension. The simplest higher-dimensional wavelet basis can be obtained via direct products of $1$-dim wavelet bases. To define them, we first label the 1-dim wavelet functions as
\begin{eqnarray}
W^\upsilon_n(z)=\left\{\begin{array}{cc}C(z^{2^{n-1}})\prod_{j=1}^{n-1}C(z^{2^{j-1}}),&~\upsilon=1\\
D(z^{2^{n-1}})\prod_{j=1}^{n-1}C(z^{2^{j-1}}),&~\upsilon=2\end{array}\right.
\label{wavelet1D}
\end{eqnarray}
so that $\upsilon=1,2$ corresponding to the IR and UV wavelets in layer $n$, respectively. The $D$-dimensional wavelet functions can then be defined by
\begin{equation}
W^{\upsilon_1\upsilon_2...\upsilon_D}_n(z_1,...,z_D)=\prod_{j=1}^D W^{\upsilon_j}_n(z_j)
\label{WD}
\end{equation}
These $2^D$ wavefunctions for $\upsilon_j=1,2$ include one IR wavelet defined by $\upsilon_j=1,~\forall j$ and $2^D-1$ other wavelets that are regarded as UV degrees of freedom. The bulk correlators $C^{\mu\nu}$ can be obtained from the boundary correlator $G_{\vec q}(\tau)$ via this basis transform:
\begin{eqnarray}
&&C^{\mu\nu}(n_1,n_2, \Delta(2^n\bold x),\tau)\notag\\&=&\sum_{\bold q} W^{\mu*}_{n_1}(\bold q)W^{\nu}_{n_2}(\bold q)e^{i \bold q \cdot \Delta (2^n\bold x) }G_{\bold q }(\tau)\label{bulkcorrelator}
\end{eqnarray}
where we have denote the $2^D-1$ dimensional label of UV states $\upsilon_1\upsilon_2...\upsilon_D$ (with at least one $\upsilon_j=2$) by $\mu$ or $\nu$ for simplicity. 

When there is $\aleph$ number of orbitals at each site, the correlation matrix is $(2^D-1)\aleph \times (2^D-1)\aleph$. For analyzing the asymptotic bulk geometry, it suffices to consider only the slowest decaying elements of $C^{\mu\nu}$, which is determined by the lowest power of ${\bf q}$ in $W^\mu_n({\bf q})$. Using the long wavelength asymptotic behavior $C(e^{iq_j})\approx \sqrt{2}$ and $D(e^{iq_j})\approx \frac{-iq_j}{\sqrt{2}}$, we see that among the $2^D-1$ UV wavelets $W^{\upsilon_1\upsilon_2...\upsilon_D}_n({\bf q})$, the ones that play a leading role in the long wavelength correlation functions are those with only one $\upsilon_j=2$ and all other $\upsilon_k=1,~k\neq j$. Therefore the asymptotic behavior of the bulk correlator is given by  
\begin{eqnarray}
C^{\mu\nu}(n,n, \Delta\bold X,\tau)\sim 2^{D(n-2)}\partial_{X_{j}}\partial_{X_{k}}\sum_{\bold q} e^{i \bold q \cdot \Delta \bold X}G_{\bold q }(\tau)\notag\\
\label{multidim1}
\end{eqnarray}
where $X_j=2^nx_j$, and $j,k$ are the directions where a 1-dim UV wavelet function $D$ is taken, {\it i.e.}, $\upsilon_{j,k}=2$. 

In the following, we shall examine the bulk geometries of various critical boundary systems at both zero and nonzero temperature, and highlight how they are different from those of $(1+1)$-dim boundary systems.

\subsection{Critical boundary model at zero $T$}

Here, we shall examine in detail the decay properties of the bulk mutual information and bulk imaginary time correlator corresponding to a critical $(D+1)$-d boundary system. We will find that they describe a bulk geometry of a higher-dimensional AdS space, in close analogy to the $(1+1)$-dim case described previously. Similar to one-dimensional case, we consider the Dirac model in (D+1)-dimensions\cite{golterman1993,creutz2001,qi2008}:
\begin{eqnarray}
H&=&\sum_kc_k^\dagger\left[\sum_{i=1}^D\Gamma^i\sin k_i+\left(M+D-\sum_{i=1}^D\cos k_i\right)\Gamma^0\right]c_k\notag\\
\label{dirac}
\end{eqnarray}
where $\Gamma^0,\Gamma^i$ are Hermitian Dirac matrices satisfying $\left\{\Gamma^\mu,\Gamma^\nu\right\}=\delta^{\mu\nu}$ for $\mu,\nu=0,1,...,D$. 
For $M$ close to $0$, the lowest energy excitations of this system are centered around $k=0$, where the EHM defined by wavelets in Eq. (\ref{wavelet1D}) and (\ref{WD}) correctly separates low energy and high energy degrees of freedom. In the following, we study the behaviors of the correlation function and dual geometry along different directions. For simplicity, we shall only explicitly study the $(2+1)$-dim case.
\\
\noindent{\bf Spatial (angular) directions.} 

We build on the results of the decay of unequal spin propagators $u,v$ for $(1+1)$-dim, with two obvious extensions mandated by Eq. \ref{multidim1}: Firstly, we now need to perform a multi-dimensional sum over $\bold q$ and secondly, we need to decide which sequence of derivatives $\partial^2_{X_{j_i}}$ produce the slowest decaying correlator.

We recall that in the absence of a EHM transform, the critical correlator behaves like the inverse first power of distance, i.e. $\sim \frac{1}{\sqrt{x^2+y^2}}$. Under the EHM transform, the correlators $u$ or $v$ decay faster due to derivatives introduced by UV projectors $D(e^{iq})$. Their slowest-decaying elements involve $D=2$ derivatives, i.e. $\partial^2_X$, $\partial^2_Y$ or $\partial^2_{XY}$. Hence  
\begin{equation} |u|\sim |v|\sim \frac{1}{(x^2+y^2)^{3/2}} \end{equation}
i.e.
\begin{equation} I_n(\Delta \bold x)\sim 6 \log |\Delta \bold x|+\text{const.} \label{Ixyangular2d}\end{equation}
which is an almost trivial generalization of the result in $(1+1)$-dim (Eq. \ref{Ixyangular}). The undetermined constant defines the AdS radius of the corresponding AdS geometry, and is a complicated function of the full correlator involving $u\sim v$. Here, we have not been careful in keeping track of the powers of $2^n$, and readers interested in doing so are invited to generalize the more rigorous derivation in Appendix \ref{app:critical}. The apparent isotropy of Eq. \ref{Ixyangular2d} may not be exact due to the numerous approximations made. However, any angular dependency should only manifest itself as a form factor in the correlators, with the leading $\log$ term in the mutual information remaining unaffected.
\\
\noindent{\bf Imaginary time direction.}

A critical $D+1$-d boundary system also has power-law decaying imaginary time correlators, consistent with the interpretation of the bulk as a higher-dimensional AdS spacetime. Explicitly,
\begin{equation}
\text{Tr } C_n(\tau)\sim \left(\frac{2^{n-1}}{v_F \tau}\right)^{D+2}
\label{multidim4}
\end{equation}
with dimension-dependent critical exponent of $D+2$. This is unlike that of the spatial correlators, which do not depend on $D$. Interpreted as a bulk geodesic distance $d_\tau$, we have
\begin{equation}
\frac{d_\tau}{\xi_\tau}=-\log \text{Tr } C_n(\tau)\sim (D+2) \left[\log(v_F\tau)-n\log 2\right] + \text{const.}
\end{equation}
which is proportional to the dimensionality. Physically, we can understand the origin of the $D+2$ exponent as follows. Each spatial direction provides an additional dimension for the decay, and contributes a power of $\frac{2^n}{v_F \tau}\propto \frac{L}{v_F \rho \tau}$. There has to be at least one direction where only the UV half of the degrees of freedom are selected, since a separation of energy scales is necessary for the EHM network. This direction contributes an additional power of $2$ due to the gradient-like property of the UV projector $D(z)$. The above statements are justified with more mathematical rigor in Appendix \ref{app:multidimtime}, where the subleading powers in the decay are also explicitly evaluated.

\subsection{Boundary model with generic dispersion at nonzero temperature}

When an energy scale is introduced by the temperature $T$, the decay of correlators is dominated by the energy scale which is independent of the EHM basis. Hence we can calculate the bulk correlators in a way similar to that of the $(1+1)$-dim case, taking note only of the multidimensionality of $\bold q$.

Instead of just analyzing the Dirac model, we make our discussion more general by allowing our energy dispersion to take the following generic asymptotic form:
\begin{equation}
E_{\bold q}= \sqrt{\sum_j^D v_j^2 q_j^{2\gamma_j}}
\label{criticaldispersion}
\end{equation}
This form encompasses various physical scenarios, and reduces to the linear Dirac model in the simplest case of $\gamma_j=1$. When $D=1$ and $\gamma>1$, Eq. \ref{criticaldispersion} represents the nonlinear dispersions discussed previously. More interestingly, it can also describe semi-Dirac points characterized by anisotropic dispersions, i.e. with $D=2$, $\gamma_1=1$ and $\gamma_2=2$. Such dispersions have been observed in realistic systems involving ultrathin (001) $VO_2$ layers embedded in $TiO_2$, which exhibit unusual electromagnetic properties\cite{pardo2009, banerjee2009,delplace2010}.

According to Eq. \ref{decay} and subsection \ref{sec:temp}, the correlators and hence mutual information decay exponentially according to the complex root of $\cosh\frac{\beta E_{\bold q}}{2}=0$ closest to the real axis. The roots are given by the values of $\bf{q}$ satisfying
\[E_{\bold q}^2 = -\pi^2 T^2 (2l+1)^2,\]
with $l\in \mathbb{Z}$, with the temperature $T$ functioning as an imaginary gap. To find the exponential decay rate in direction $j$, we have to find 
$|Im(q_j)|$, the imaginary part of the complex root $q=q_j$ of
\begin{equation}
v_j^2 q^{2\gamma_j}=-(m^2+\pi^2 T^2 (2l+1)^2)
\label{criticalTD}
\end{equation}
where $m^2=\sum_{i\neq j}^D v_i^2 q_i^{2\gamma_i}$ denotes an effective mass from the momentum contributions from all the other directions. Since $T^2$ and $m^2$ are both positive, we clearly choose $l=0$ for $q_j$ to have the smallest imaginary part, i.e. slowest decay. While the decay rate also depends on $m$, the combination of momentum components giving $m=0$ yields the slowest decay rate $|Im(q_j)|$. We can take $m=0$ to be the dominant contribution to the overall decay rate $h_j$, and the $m>0$ contributions as the subleading corrections. This will be discussed explicitly for the two cases below, with calculational details relegated to Appendix \ref{app:temp2}.

\subsubsection{Finite temperature Dirac fermions}

We first discuss the massless Dirac case with
\begin{equation}
E_{\bold q}= v|\bold q| = v\sqrt{\sum_j^D  q_j^{2}},
\label{multidirac}
\end{equation}
i.e. with $v_j=v$ and $\gamma_j=1$ for $j=1,2,...,D$. Let $\Delta \vec x$ be the displacement between two distant points within the same layer in the bulk. The mutual information decays like $I_{xy}\sim 8|u|^2\sim 8|v|^2$ where, as shown in Appendix \ref{app:temp2},
\begin{eqnarray}
u\sim v &\sim& \int e^{i2^n \bold q \cdot \Delta \vec x}\tanh\frac{\beta E_{\bold q}}{2}d^D \bold q \notag\\
&\sim& e^{-\frac{2^n\pi T}{v}|\Delta \vec x|}
\label{multidirac2}
\end{eqnarray}
Hence
\begin{equation}
-\log I_n( \Delta \vec x)\sim 	\frac{2^{n+1}\pi T}{v}|\Delta \vec x| =\frac{LT}{v} \sqrt{\sum_j^D (\Delta \theta_j)^2}\\
\label{multidirac3}
\end{equation}
This is a direct generalization of Eq. \ref{temp1} for the $(1+1)$-dim  critical Dirac model at temperature $T$, whose bulk geometry corresponds to that of a BTZ black hole horizon in the IR limit. Here, we have exactly the same asymptotic behavior, with the horizon having the same topology as the boundary system. When the latter is defined on a D-dimensional lattice with periodic boundary condition along all directions, the horizon is a D-dimensional torus $T^D$.

\subsubsection{$(2+1)$-dimensional anisotropic dispersion}

We now consider a generic anisotropic dispersion in $D=2$, and show that the event horizon can also become anisotropic. The dispersion is given by
\begin{equation}
E_{\bold q}=  \sqrt{ v_1^2 q_1^{2\gamma_1}+ v_2^2 q_2^{2\gamma_2}},
\label{multianisotropic1}
\end{equation}
with correlator decay rates $\left(\frac{\pi^2 T^2 +v_{j'}^2q_{j'}^2}{v_j^2}\right)^{\frac{1}{2\gamma_j}}\sin \frac{\pi}{2\gamma_j}$ where $j'=1,2$ for $j=2,1$. It is mathematically tricky to obtain the asymptotic behavior of $I_{xy}$ for arbitrary $\Delta \vec x$, when all components of $x$ are not small. For our current purpose, it suffices to expand the asymptotic behavior about the limiting directions $\Delta \vec x = x \hat e_x$ and $ y\hat e_y$. After a Gaussian integral computation detailed in Appendix \ref{app:temp2}, the mutual information at $\Delta \vec x = |\Delta \vec x|(\cos\phi  \hat e_x + \sin\phi  \hat e_y)$ for $\phi$ near $0$ is approximately given by
\begin{eqnarray}
&&-\log I_n(\vec \Delta x )|_{\phi  \approx 0}\notag\\
&\sim& 	2^{n+1}\left (\sin\frac{\pi}{2\gamma_1}\left(\frac{\pi T}{v_1}\right)^{1/\gamma_1}x+\frac{\gamma_1(\pi T)^{2-1/\gamma_1}}{2\sin\frac{\pi}{2\gamma_1}}\frac{v_1^{1/\gamma_1}}{v_2^2}\frac{y^2}{x}\right)\notag\\
&=& 2^{n+1}|\Delta \vec x|\sin\frac{\pi}{2\gamma_1}\left(\frac{\pi T}{v_1}\right)^{1/\gamma_1}\left(1+\frac{\left(\alpha_1^2-1\right)}{2}\phi ^2 + O\left(\phi ^4\right)\right)\notag\\
\label{multianisotropic2}
\end{eqnarray}
where $\alpha_j=\sqrt{\gamma_j}\frac{(\pi T)^{1-1/\gamma_j}v_1^{1/\gamma_j}}{v_{\bar j}\sin \frac{\pi}{2\gamma_j}}$. An exactly analogous result holds near $\phi  \approx \frac{\pi}{2}$, with $\gamma_1,\alpha_1$ and $v_1$ replaced by $\gamma_2,\alpha_2$ and $v_2$. 

Notably, in the isotropic linear Dirac case where $\gamma_j=1$ and $v_1=v_2$, $\alpha_j=1$, the mutual information is manifestly asymptotically isotropic to third order by Eq. \ref{multianisotropic2}. This is despite the fact that the  wavelet basis was constructed via tensor products of those of each direction and hence only possess four-fold rotation symmetry. 
\begin{figure}[H]
\centering
\includegraphics[scale=.99]{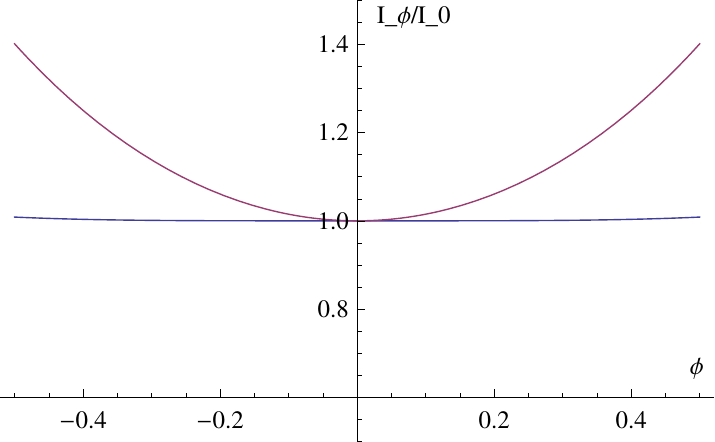}
\caption{Plots of $I_{xy}(\phi)/I_{xy}(\phi=0)$ for $\alpha=1$ (almost horizontal line) and $\alpha=2$ (upper curved line) according to Eq. \ref{multianisotropic2}, with higher order terms in $\phi^2$ kept. We see that in the isotropic linear Dirac case with $\alpha=1$, $I_{xy} \propto |\Delta \vec x|$ to a high degree of accuracy even away from $\phi\ll 1$. }\label{horizon_angle} \end{figure}

By contrast, when the dispersion acquires some nonlinearity, $\gamma_j>1$ and $\alpha_j\neq 1$ for some $j$ and we expect the mutual information to be significantly anisotropic in a temperature dependent way. This is illustrated in Fig. \ref{horizon_angle}, where the angular dependence of the mutual information is compared for $\alpha=1$ and $2$. In any case, the factor $2^{n+1}\Delta |\vec{x}|$ in Eq. (\ref{multianisotropic2}) suggests that there is still a finite area (anisotropic) horizon in the IR limit, since the circumference of a closed circle around any periodic direction approaches a finite value while $n\rightarrow \infty$.

\section{Conclusion}
\label{sec:concl}

In this work, we have analytically studied the emergent bulk geometries of several different boundary systems through the EHM approach. In general, critical boundary systems at zero temperature correspond to scale invariant bulk geometries. A spatial boundary appears in the infrared region when a mass scale is introduced by a nonzero mass. At nonzero temperature, a horizon appears in the infrared region, which is distinguished from the spatial boundary by the infinite red-shift that can be observed in the behavior of correlation functions along the imaginary time direction. For critical boundary theories with different dynamical exponents in time spatial and temporal directions, the spatial geometry is similar but the space-time geometry depends on the dynamical exponent, at both zero temperature and nonzero temperature. Most of the above results qualitatively still hold true when the EHM is generalized to higher dimensions.

A major open question concerns the dual geometry for a Fermi gas with finite charge density. The existence of a nonzero Fermi momentum makes it inappropriate to use the same EHM mapping defined here, since the long wavelength limit will no longer correspond to the low energy limit. A modified tensor network is required in order to describe the correct infrared physics near the Fermi surface. 


\begin{acknowledgements}
We thank Chao-Ming Jian and Yingfei Gu for helpful discussions. 
XLQ is supported by the David and Lucile Packard foundation.
\end{acknowledgements}

\bibliography{ehm,entanglement}

\newpage
\appendix

\section{Properties of the Mutual Information}

\subsection{Relation to the single-particle correlators}
\label{app:mutualcorr}

The mutual information as given by Eq. \ref{mutual} contains the two-site entanglement entropy, which depends on the two-site correlator $C_{xy}$ via Eq. \ref{Ixy00}. 

It is a matrix of single-particle propagators $[C_{xy}]_{ij}=\langle\beta_i \beta_j^\dagger\rangle$, where $i,j\in \{ x,y\}$ and $\b_x,\b_y$ are the bulk annihilation operators. Below, we shall write
\begin{equation}C_{xy}=\left(\begin{matrix}
 & C_x & C_{x-y} \\
 & C_{x-y}^\dagger  & C_y \\
\end{matrix}\right)=C_0 + V
\label{Cxy}\end{equation}
where $C_x$ and $C_y$ the on-site single-particle correlators, and $C_{x-y}$ is the (non-hermitian) single-particle propagator from site $y$ to $x$. The off-diagonal contribution $C_{x-y}$ decays rapidly for large $|x-y|$, and $I_{xy}$ can be accurately approximated in that limit. Below, we write $C_{xy}=C_0 + V$, where $C_0=C_x\otimes C_y$ and $V$ is a perturbation containing the off-diagonal parts $C_{x-y}$ and $C_{x-y}^\dagger$:
\begin{widetext}
\begin{eqnarray}
I_{xy}&=&S_x+S_y-S_{xy}\notag\\
&=& S_x+S_y+\text{Tr }((C_0+V)\log(C_0+V)+(1-C_0-V)\log(1-C_0-V))\notag\\
&\approx & S_x+S_y+\text{Tr }((C_0+V)(\log C_0 + C_0^{-1}V-\frac{1}{2}(C_0^{-1}V)^2)\notag\\&&+(1-C_0-V)(\log (1-C_0) + (1-C_0)^{-1}V-\frac{1}{2}((1-C_0)^{-1}V)^2))\notag\\
&\approx& (S_x+S_y-\text{Tr }(S_x\otimes S_y))+2 \text{Tr } V + \text{Tr } [V (\log C_0 + \log(1-C_0))]+\frac{1}{2}\text{Tr } VC_0^{-1}V  +\frac{1}{2}\text{Tr } V(1-C_0)^{-1}V  \notag\\
&=& (S_x+S_y-S_x-S_y)+\frac{1}{2}\text{Tr } VC_0^{-1}V  +\frac{1}{2}\text{Tr } V(1-C_0)^{-1}V  \notag\\
&\approx& \frac{1}{2}\text{Tr }[ V(C_0(1-C_0))^{-1}V  ]\notag\\
&\approx& \frac{1}{2}\text{Tr }\left[C_{x-y}\frac{1}{C_y(\mathbb{I}-C_y)}C_{y-x}+(x\leftrightarrow y)\right]\notag\\
&\sim & \text{Tr } [C^\dagger_{y-x}C_{y-x}]
\label{Ixy0}
\end{eqnarray}
\end{widetext}
where $C_x,C_y$ are the single-particle onsite correlators, and $C_{x-y}$ is the single-particle propagator between the two different sites $x$ and $y$. Note that we have implicitly assumed that $C_0$ and $V$ commute while going from lines 2 to 3, which holds in the IR limit.

\subsection{The mutual information in terms of the reduced density matrix}
\label{app:mutual}

The mutual information can also be understood as the Kullback-Liebler divergence\cite{cover1991} of the distributions described by $\rho_{\bold x \bold y}$ and $\rho_{\bold x}\rho_{\bold y}$, where $\rho_{\bold x}$ is the single-site reduced-density matrix (RDM) and $\rho_{\bold x \bold y}$ is the two-site RDM. Writing the trace over one site as $\text{Tr }_{\bold x}$ and two sites as $\text{Tr }_{\bold x \bold y}$, we have
\begin{eqnarray}
I_{\bold x\bold y} &=& S_{\bold x} + S_{\bold y} - S_{\bold x \bold y}\notag\\
&=& - \text{Tr }_\bold y \rho_\bold x\log \rho_\bold x- \text{Tr }_\bold x \rho_\bold y\log \rho_\bold y + \text{Tr }_{\bold x \bold y} \rho_{\bold x \bold y}\log \rho_{\bold x \bold y}\notag\\
&=& - \text{Tr }_{\bold x \bold y} \rho_{\bold x \bold y}\log \rho_{\bold x }-\text{Tr }_{\bold x\bold y } \rho_{\bold y}\log \rho_{\bold y} + \text{Tr }_{\bold x \bold y} \rho_{\bold x \bold y}\log \rho_{\bold x \bold y}\notag\\
&=& \text{Tr }_{\bold x \bold y}\rho_{\bold x \bold y}\log \frac{\rho_{\bold x \bold y}}{\rho_\bold x \rho_\bold y}\notag\\
&=& \mathbb{E}\left[\log \frac{\rho_{\bold x \bold y}}{\rho_\bold x \rho_\bold y} \right]
\label{mutual2}
\end{eqnarray}
To maximize the mutual information $I_{\bold x \bold y}$, we need $\rho_{\bold x \bold y}$ and $\rho_\bold x \rho_\bold y $ to be as different as possible. Since the $\log$ function drops steeply below unity, we especially want to avoid situations where the (square root of the) single-site RDM eigenvalue is large compared to that of the two-site RDM. Since $\rho_\bold x = \text{Tr }_\bold y \rho_{\bold x \bold y}$, the above-mentioned situation is more likely when the RHS contains a large number of contributions from states belonging to different $y$. Hence we conclude that a maximal $I_{\bold x \bold y}$ must have minimal spread of $\rho_{\bold x \bold y}$, i.e. have 2-particle states that are maximally entangled in the precise sense of Eq. \ref{mutual2}. Note that this scenario represents a hypothetical optimum, and may not be realized the physical systems that we have discussed.

\section{Correlators for two-band fermionic systems with particle-hole symmetry}
\label{app:twoband}

Here we derive the detailed results for two-band, particle-hole symmetric models studied in this work. They hold for generic two-band models at arbitrary chemical potential $\mu$, although we will ultimately use them only for the Dirac model at $\mu=0$.

Due to particle-hole symmetry, the onsite correlator (projector) $C_x$ takes the following form in spin space:
\begin{equation}C_x=\langle \b_x \b^\dagger_x\rangle=\left(\begin{matrix}
 & a & i A \\
 & -i A  & a \\
\end{matrix}\right)\label{Cx}\end{equation}
where $a$ and $A$ are real, up to inconsequential corrections at nonzero temperatures. At zero chemical potential $\mu$, we always have $a=\frac{1}{2}$. The single-particle propagator is slightly more complicated:
\begin{equation}C_{x-y}=\langle \b_x \b^\dagger_y\rangle=\left(\begin{matrix}
 & b & i u \\
 & i v  & b \\
\end{matrix}\right)\label{Cxy2}\end{equation}
When $\mu=0$, the equal-spin propagator $b=0$ and the unequal-spin propagators $u,v$ are real. At nonzero $\mu$, $u,v$ are not necessarily real but $b$ is still real. $|u|\neq|v|$ in general, since $C_{x-y}$ does not have to be hermitian in spin space alone. When the onsite propagators $C_x=C_y$, as in the case when they are related by translation symmetry in the same bulk layer, we obtain upon substituting Eqs. \ref{Cx} and \ref{Cxy2} into Eq. \ref{Ixy0}
\begin{eqnarray}
&&I_{xy}=\notag\\
&& \frac{[a(1-a)-A^2][2b^2 +(|u|^2+|v|^2)]+2bA(1-2a)Re[u-v]}{(A^2-a^2)(A^2-(1-a)^2)}\notag\\
\label{Ixy}
\end{eqnarray}
where $a,A,b,u,v$ are defined in Eqs. \ref{Cx},\ref{Cxy2} and \ref{Cxy}. At zero chemical potential, $b=0$ and $a=\frac{1}{2}$, and Eq. \ref{Ixy} simplifies further to
\begin{equation}
I_{xy}|_{\mu=0}=\frac{4(|u|^2+|v|^2)}{1-4A^2}
\label{Ixy2}
\end{equation}
Since the onsite contribution $A$ does not vary with the spatial displacement $\bold x -\bold y$, the decay properties of $I_{xy}$ at $\mu=0$ depends almost entirely on $u$ and $v$.

From Eq. \ref{Cx}, it is almost trivial to write down the expression of the single-site Entanglement entropy (EE):
\begin{eqnarray}
S_x&=&-\text{Tr }(C_x \log C_x+ (\mathbb{I}-C_x)\log(\mathbb{I}-C_x))\notag\\
&=&-2\sum_{\lambda_\pm}\left[\lambda_\pm \log \lambda_\pm+(1-\lambda_\pm)\log(1-\lambda_\pm)\right]\notag\\
\label{entropy3}
\end{eqnarray}
where $\lambda_\pm=a\pm |A|$ are the eigenvalues of $C_x$. When $\mu=0$, $a=\frac{1}{2}$ and the EE is maximal at $S_x^{max}=\log 4$ at $A=0$. Indeed, without unequal-spin correlation ($A=0$), we have no information about what is happening to the other spin state. When $A$ is small, the maximal entropy is corrected according to $S_x\approx S ^{max}_x-2A^2=\log 4 -2A^2$. $S_x$ vanishes when $a=A=\frac{1}{2}$, which produces a pure eigenstate $\propto |-\rangle+|+\rangle$.

We now present the explicit forms of the correlators, denoted collectively as $G_{\bf{q}}$. Given a Hamiltonian $h_\bold q$, $G_\bold q(\tau)=e^{\tau(h_{\bold q}-\mu )}(\mathbb{I}+e^{\beta(h_{\bold q}-\mu)})^{-1}
$ (Eq. \ref{gq0}) is explicitly
\begin{widetext}
\begin{equation}
G_\bold q(\tau)=\frac{e^{-\tau \mu}}{2}\left(\cosh(\tau E_\bold q)\mathbb{I}+\frac{h_\bold q}{E_\bold q}\sinh(\tau E_\bold q)\right)\left(\left(1+\frac{\sinh \beta \mu}{\cosh \beta E_\bold q+\cosh \beta \mu}\right)\mathbb{I}-\frac{h_\bold q}{E_\bold q}\frac{\sinh\beta E_\bold q}{\cosh \beta E_\bold q +\cosh \beta \mu}\right)\\
\label{gq}
\end{equation}\end{widetext}
In the limit of zero temperature $\beta\rightarrow \infty$, the equal-time correlator $G_\bold q=G_\bold q (\tau=0)$ tends to
\begin{eqnarray} G_\bold q&\rightarrow& G_\bold q|_{\mu=0}+ \frac{1}{2}\theta(|\mu|-E_\bold q)\left(\text{sgn}(\mu)\mathbb{I}+\frac{h_\bold q}{E_\bold q} \right)\notag\\
&=&\theta(\mu-E_\bold q)\mathbb{I}+\theta(E_\bold q-|\mu|)\frac{1}{2}\left(\text{sgn}(\mu)\mathbb{I}-\frac{h_\bold q}{E_\bold q}\right)\notag\\
\label{mueq}
\end{eqnarray}
The physical interpretation of the above  is clear: When $E_\bold q>|\mu|$, $G_\bold q$ is exactly the same as in the $\mu=0$ case. For $E_\bold q<|\mu|$, $G_\bold q$ projects identically to either both bands or none depending on the sign of $\mu$.

For zero $\mu$ but nonzero temperature, we have
\begin{equation}
G_\bold q|_{\mu=0}=\frac{1}{2}\mathbb{I}-\frac{h_\bold q}{2E_\bold q}\tanh\frac{\beta E_\bold q}{2}
\label{gqtemp}
\end{equation}
Of course, This further reduces to the usual projection operator given by $\frac{1}{2}\left( \mathbb{I}-\frac{h_\bold q}{E_\bold q}\right)$ at zero-temperature.

\section{Derivation of the bulk mutual information for $(1+1)$-dim  critical boundary systems at $T=0$}
\label{app:critical}
We start from the following expression for the mutual information $I_{xy}$ between sites $x$ and $y$ (Eq. \ref{Ixy2}):
\begin{equation}
I_{xy}|_{\mu=0}=\frac{4(|u|^2+|v|^2)}{1-4A^2}\sim 8|u|^2\sim 8|v|^2
\end{equation}
where, as introduced in the main text and the previous appendix, $u$ and $v$ are the unequal-spin single particle propagators, and $A$ the unequal-spin onsite propagator which tends to zero beyond moderately large $n$. Below, we shall derive their asymptotic behavior in detail.

\subsection{Angular direction}
\label{app:critical1}

We evaluate Eq. \ref{uv} for a large angular interval of $\Delta x$ sites by deforming the contour around the branch cut from $z=0$ to $z=\infty$. (looking like a tight-lipped Pac-man):
\begin{eqnarray}
u\sim v &\sim&\frac{1}{2}\int_0^1 W_n^*(z^{-1})W_n(z)z^{2^n\Delta x}(\sqrt{z}-\sqrt{e^{2\pi i}z})\frac{dz}{z}\notag\\
&=&\int_0^1 W_n^*(z^{-1})W_n(z)z^{2^n\Delta x}\frac{1}{\sqrt{z}}dz\notag\\
&=&\int_0^1 (W_n^*(z^{-1})W_n(z)z^{2^n})z^{2^n(\Delta x-1)-1/2}dz\notag\\
&=&\int_0^1 Q(z)z^X dz
\label{u2}
\end{eqnarray}
where $W_{n}(z)=\frac{1}{\sqrt{2\pi}}D(z^{2^{n-1}})\prod_{j=1}^{n-1}C(z^{2^{j-1}})$ and $C(z),D(z)=\frac{1\pm z}{\sqrt{2}}$. We have decomposed the integrand into a term $Q(z)=W_n^*(z^{-1})W_n(z)z^{2^n}$ that does not have negative powers of $z$, and $z^X$ with $X=2^n(\Delta x-1)-1/2$ still very large. We next integrate by parts to get the asymptotic behavior of $u\sim v$:
\begin{eqnarray}
u\sim v&\sim &\int_0^1 Q(z)z^X dz\notag\\
&=&\frac{Q(1)}{X+1}-\frac{1}{X+1}\int_0^1Q'(z)z^{X+1}dz\notag\\
&=& \frac{Q(1)}{X+1}-\frac{Q'(1)}{(X+1)(X+2)}+...\notag\\
&=& 0-0 + \frac{Q''(1)}{(X+1)(X+2)(X+3)}+...\notag\\
&\sim & \frac{Q''(1)}{X^3}
\label{u3}
\end{eqnarray}
Here, we have stopped at the 2nd derivative of $Q$, because it is the lowest nonzero derivative at $z=1$. $Q(1)=Q'(1)=0$ because they each must contain at least one factor of $W_n(1)$ or $W_n^*(1)$, both of which are zero due to the presence of UV projectors $D(z^{2^{n-1}})$. We also truncate off higher derivative terms as they contain higher powers of $\frac{1}{X}$. As one may expect, $u$ or $v$ depends \emph{exclusively} on the behavior of the EHM basis at $z=1$ or $q=-i\log z = 0$, the IR point where criticality occurs.

Substituting the explicit form of $Q(z)$ and differentiating, we obtain
\begin{eqnarray}
u\sim v&\sim & \frac{Q''(1)}{X^3}\notag\\
&=& \frac{2^{2(n-1)}}{\pi}|C(1)^{n-1}D'(1)|^2\frac{1}{X^3}\notag\\
&=& \frac{2^{2(n-1)}}{\pi}\left|\sqrt{2}^{n-1}\left(-\frac{1}{\sqrt{2}}\right)\right|^2\frac{1}{X^3}\notag\\
&=& \frac{2^{3n-4}}{\pi}\frac{1}{(2^n\Delta x)^3}\notag\\
&=& \frac{1}{16\pi}\frac{1}{(\Delta x)^3}
\label{u4}
\end{eqnarray}
That $n$ drops out is not a coincidence, but a manifestation of scale invariance. Note that the presence of the critical point, which is a property of the Hamiltonian, only ensures that the first line of Eq. \ref{u2} will not evaluate to zero; the power-law decay rate is \emph{entirely} determined by the analytic properties of the chosen EHM basis at that IR point. In this case, the mutual information behaves asymptotically as
\begin{equation}
I_{xy}=I_n(\Delta x)\sim\frac{1}{32\pi^2}\frac{1}{(\Delta x)^6}
\end{equation}

\subsection{Radial direction}
\label{app:critical2}

Here, we evaluate Eq. \ref{uv2} for the mutual information between sites that are separated radially. When one of the layer lies at the UV boundary of the bulk (layer $1$), $u$ and $v$ between layers $1$ and $n$ can be easily evaluated viz.
\begin{eqnarray}
u,v&=&-i\oint_{|z|=1}\frac{dz}{z}W^*_{1}(z^{-1})W_{n}(z)z^{\mp \frac{1}{2}}\notag\\
&=&-i\oint_{|z|=1}\frac{dz}{z}\frac{1}{2\pi\sqrt{2^{n+1}}}(1-z^{-1})\frac{(1-z^{2^{n-1}})^2}{1-z}z^{\mp \frac{1}{2}}\notag\\
&=&i\oint_{|z|=1}\frac{dz}{z^{2\pm \frac{1}{2}}}(1-2z^{2^{n-1}}+z^{2^n})\notag\\
&=&i\oint_{|z|=1}\frac{dz}{z^{2\pm \frac{1}{2}}}-i\oint_{|z|=1}\frac{(2z^{2^{n-1}}-z^{2^n})dz}{z^{2\pm \frac{1}{2}}}\notag\\
\end{eqnarray}
The first integrand diverges when $z\rightarrow 0$, so the contour should be inverted about $|z|=1$ and closed at infinity. The second one diverges when $z\rightarrow \infty$ for $n\geq 2$, so the contour should be closed like a Pac-man.

Performing the resultant real integrals analogously to Eq. \ref{u2}, we obtain
\begin{equation}
|u|=\frac{1}{2\pi\sqrt{2^{n-1}}}
\end{equation}
\begin{equation}
|v|=\frac{1}{6\pi\sqrt{2^{n-1}}}
\end{equation}
The single-particle bulk propagator is not hermitian in spin-space, with $u\neq v$. The above are exact results, not asymptotic ones. However, exact results like these usually do not exist for more generic wavelet mappings. The resultant mutual information is
\begin{equation}
I_{xy}=I(1,n)\sim 4(|u|^2+|v|^2)=\frac{10}{9\pi^2}\frac{1}{2^{n-1}}
\label{radialmutual}
\end{equation}

\section{Calculational details for the bulk mutual information at nonzero temperature for critical systems in arbitrary number of dimensions}
\label{app:temp}

In this appendix, we shall fill in the mathematical gaps in the derivations of various nonzero $T$ results for gapless systems. Since the mutual information $I_{xy}$ between sites $x$ and $y$ is (Eq. \ref{Ixy2}),
\begin{equation}
I_{xy}=\frac{4(|u|^2+|v|^2)}{1-4A^2}\sim 8|u|^2\sim 8|v|^2
\end{equation}
where $A$ trivially approaches zero beyond the first few layers $n$, we will just need to find the asymptotic behavior of the unequal-spin single particle propagators (two-point functions) $u\sim v$.

\subsection{Decay of nonzero $T$ correlators for generic critical dispersions}
\label{app:temp2}

We consider the energy dispersion Eq. \ref{criticaldispersion}

\begin{equation}  E_{\bold q} = \sqrt{\sum_j^D v_j^2 q_j^{2\gamma_j}} \end{equation}
In the $j^{th}$ direction, the decay rate of the correlators $u,v$  is given by the the imaginary part of the root $q=q_j$ of
\begin{equation}
v_j^2 q^{2\gamma_j} = -(m^2+\pi^2 T^2)
\end{equation}
nearest to the real axis, where $m^2=\sum_{i\neq j}^D v_i^2 q_i^{2\gamma_i}$ represents an effective mass from the momentum contributions from all the other directions. $m^2\ll T^2$ is always satisfied for a bulk layer sufficiently deep in the IR, i.e. of sufficiently large $n$. In terms of the variable $z=e^{iq}=e^{iq_j}$ or $q=q_j=\frac{z-z^{-1}}{2i}$ used previously, the decay rate is given by $-\log|z_0|$, where $z_0$ is the root of 
\begin{equation}
z^4 + 2(\alpha_j-1)z^2+1 =0
\end{equation} 
within the unit circle and closest to its boundary, with $\alpha_j=\frac{2e^{i\pi/\gamma}}{v_j^{2}}\left((\pi T)^2 + m^2\right)^{1/\gamma}$.  Solving the above equation and taking the imaginary part,
\begin{eqnarray}
|Im(q_j(m))|&=& \frac{\sin \frac{\pi}{2\gamma_j}}{v_j^{1/\gamma_j}}\left((\pi T)^2 + m^2\right)^{1/(2\gamma_j)}\notag\\
&\approx & U_j\left(1+ V_j\sum_{i\neq j}^D v_i^2 q_i^{2\gamma_i}\right)
\end{eqnarray}
where $U_j=\sin \frac{\pi}{2\gamma_j}\left(\frac{\pi T}{v_j}\right)^{\frac{1}{\gamma_j}}$ and $V_j=(2\gamma_j (\pi T)^2)^{-1}$. However, this is not the physical decay rate as it still depends on the other momenta. To obtain the physical decay rate, we integrate over the latter (taking $j=1$ without loss of generality):
\begin{eqnarray}
u\sim v &\sim& \int e^{i2^n \bold q \cdot \Delta \vec x}\tanh\frac{\beta E_{\bold q}}{2}d^D \bold q \notag\\
&\sim & \prod_{j=2}^D \int dq_j e^{-2^n |Im (q_1(m))|\Delta x_1 }e^{i2^n\sum_{j\geq 2}^D \Delta x_jq_j}\notag\\
&\approx & e^{-2^nU_1 \Delta x_1}\prod_{j=2}^D\left[\int dq e^{-(U_1V_1v_j^22^n \Delta x_1) q^2}e^{i2^n\Delta x_j q}\right]\notag\\
&\sim & e^{-2^nU_1 \Delta x_1}\prod_{j=2}^D e^{-\frac{1}{4U_1V_1 v_j^2}\frac{(2^n\Delta x_j)^2}{2^n\Delta x_1}}\notag\\
&= &  e^{-2^nU_1 \Delta x_1-\frac{2^n}{4U_1V_1 \Delta x_1}\sum_{j\geq 2}^D \frac{(\Delta x_j)^2}{v_j^2}}\notag\\
\label{genericdecay1}
\end{eqnarray}
This is just Eq.\ref{multianisotropic2} for the general anisotropic case. For the isotropic linear Dirac case where $U_j=\frac{\pi T}{v}$ and $V_j=\frac{1}{2(\pi T)^2}$ for all $j$, Eq. \ref{genericdecay1} nicely simplifies to
\begin{eqnarray}
u\sim v &\sim &  exp\left(-2^nU_1 \Delta x_1-\frac{2^n}{4U_1V_1 \Delta x_1}\sum_{j\geq 2}^D \frac{(\Delta x_j)^2}{v_j^2}\right )\notag\\
&=&exp\left(-2^n\frac{\pi T}{v} \Delta x_1-\frac{2^n}{4\frac{\pi T}{v}\frac{1}{2(\pi  T)^2} \Delta x_1}\sum_{j\geq 2}^D \frac{(\Delta x_j)^2}{v^2}\right)\notag\\
&=&e^{-2^n\frac{\pi T}{v} \Delta x_1\left(1+\frac{\sum_{j\geq 2}^D (\Delta x_j)^2}{2\Delta x_1}\right)}\notag\\
&\approx & e^{-2^n\frac{\pi T}{v}\sqrt{\sum_j (\Delta x_j)^2}}\notag\\
&= & e^{-2^n\frac{\pi T}{v}|\Delta \vec x|}
\label{genericdecay2}
\end{eqnarray}
This shows that the correlators and hence mutual information decay isotropically in $\Delta \vec x$ for the isotropic linear Dirac model, at least in the neighborhood of $\hat e_j$, $j=1,2,...,D$. The nonlinearity of Eq. \ref{genericdecay1} is further explored in the main text around Fig. \ref{horizon_angle}.

\section{Calculational details for the imaginary time correlator}
\label{app:time}
Here we shall present the full derivations of the more involved results on the imaginary time correlator $C_n(\tau)$. We shall derive the results for an arbitrary chemical potential $\mu$, so as to illustrate the interesting continuous crossover of $C_n(\tau)$ as a nonzero chemical potential is introduced to a critical system. 

For large $\tau$ and $\mu\geq 0$, Eq. \ref{gq} and \ref{mueq} simplify to
\begin{equation}
G_{E_q< \mu}(\tau)= \frac{e^{-\mu \tau}e^{\tau E_q}}{2}\left(\mathbb{I}+\frac{h_q}{E_q}\right)
\label{gqtime1a}
\end{equation}
and
\begin{equation}
G_{E_q\geq  \mu}(\tau)= \frac{e^{-\mu \tau}e^{-\tau E_q}}{2}\left(\mathbb{I}-\frac{h_q}{E_q}\right)
\label{gqtime1b}
\end{equation}
Hence the full correlator given by 
\begin{widetext}
\begin{equation}
C_{n}(\tau)=\frac{e^{-\mu\tau}}{2}\int_{-\pi}^\pi dq |W_n(e^{iq})|^2 \left[\theta(E_q-\mu)e^{-\tau E_q}\left(\mathbb{I}-\frac{h_q}{E_q}\right)+\theta(\mu-E_q)e^{\tau E_q}\left(\mathbb{I}+\frac{h_q}{E_q}\right)\right]
\label{time}
\end{equation}
\end{widetext}
where $\theta$ is the Heaviside function. We only have to care about the extreme IR region of this integral, since $e^{-E_q\tau}=e^{-v_F|q|\tau}$ decays rapidly for large $\tau$ in a critical system.

To proceed further, we only need to understand the IR behavior of $W_n(z)=D(z^{2^{n-1}})\prod_{j=1}^{n-1}C(z^{2^{j-1}})$.
The chemical potential sets an energy scale that divides two qualitatively different regimes. When the layer $n$ is below above the energy scale of $\mu$, i.e. $n<n^*$ where $2^{n^*}=\frac{2\pi v_F}{\mu}$, $W_n(e^{iq})$ is effectively dominated by a sharp peak at $q_0=\frac{2\pi}{2^n}<\mu$. However, for $n>n^*$, $W_n(e^{iq})$ is governed by its analytic behavior near the IR point $z=1$. Perturbing away from the IR point with $z=e^{i(0+\Delta q)}\approx 1-i \Delta q$,
\begin{eqnarray}
|W_n(e^{i \Delta q})|^2&\approx & |W_n(1-i \Delta)|^2\notag\\
&=&W_n^*(1+i\Delta q)W_n(1-i \Delta q)\notag\\
&\approx &\frac{1}{2\pi} \left |\left(\prod_{j=1}^{n-1}C(1)\right)\right|^2\left |D'(1)\right|^22^{2(n-1)}(\Delta q)^{2}\notag\\
&=&\frac{2^{3n}}{32\pi} (\Delta q)^{2}
\label{u6}
\end{eqnarray}
Noting that the even part of $\frac{h_q}{E_q}$ is $\frac{|q|}{2}$, the correlator simplifies to (with $E_\nu=\mu$)
\begin{widetext}
\begin{equation}
C_{n}(\tau)\approx\frac{e^{-\mu\tau}2^{3n}}{32 \pi}
\int_{0}^\pi dq  q^{2}\left[\theta( q-\nu)e^{-\tau v_F q}\left(\mathbb{I}-\frac{q}{2}\sigma_2\right)+\theta(\nu- q)e^{\tau v_F q}\left(\mathbb{I}+\frac{ q}{2}\sigma_2\right)\right]\\
\label{time2}
\end{equation}
\end{widetext}
This integral can be exactly solved in terms of incomplete Gamma functions. However, we just want to extract the relevant asymptotic behavior set by the scale $\mu \tau$. Since the correlator captures the IR behavior, it should remain invariant even if the upper limit of $\pi$ is replaced by an arbitrary $q_{cutoff}$ in the first term on the RHS. The following formulae come in handy: $\int_0^{\mu}q^\gamma e^{q\tau}dq \sim \frac{\mu^{\gamma+1}}{\gamma+1}$ for  $\mu\tau\ll 1$ and $\frac{e^{\mu \tau}\mu^\gamma}{\tau}$ for $\mu\tau \gg 1$. Also, $\int_\mu^{q_{cutoff}}q^\gamma e^{-q\tau}dq\sim\frac{\gamma!}{\tau^{\gamma+1}}$ for $\mu\tau\ll 1$ and $\frac{e^{-\mu \tau}\mu^\gamma}{\tau}$ for $\mu\tau\gg 1$.

For the case of chemical potential discussed in the main text, $\mu\tau=0$ and we always have
\begin{eqnarray}
C_n(\tau)|_{\mu=0}&\approx&\frac{e^{-\mu\tau}2^{3n}}{16 \pi} \frac{1}{v_F^3\tau^3}\left(\mathbb{I}+\frac{3}{2v_F\tau}\sigma_2\right)|_{\mu=0}\notag\\
&\sim &\frac{1}{16 \pi}\left(\frac{2^n}{v_F\tau}\right)^3\mathbb{I}
\label{time3}
\end{eqnarray}
The extension of this result to nonlinear dispersions will be discussed in Appendix \ref{genericdirac}. Eq. \ref{time3} may also be used for the short-time behavior when $\mu\neq 0$, as long as $\mu\tau\ll 1$.

\section{Extension to the critical $(1+1)$-dim Dirac Model with $M\neq 1$ and general discussion of criticality }
\label{genericdirac}

In the main text, we have focused on the $M=1$ case of the critical (gapless) Dirac model 

\begin{equation} H_{Dirac}(k)=v_F[\sin k \sigma_1 + M(1-\cos k )\sigma_2]\end{equation}
where $M$ controls the relative weight between the $\sin k$ and $1-\cos k $ terms. These two terms have respectively linear and quadratic dispersions for small $|k|$, and here we study the effects of their interplay.

\begin{figure}[H]
\centering
\includegraphics[scale=1.5]{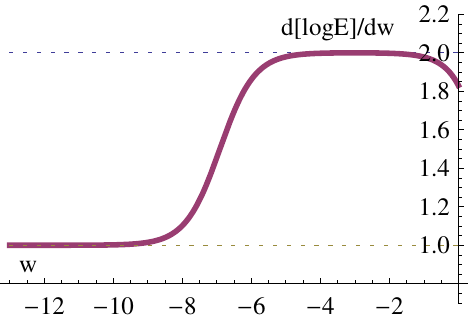}
\caption{$\frac{d\log E_w}{dw}$ for $M=2000$. We see that the dispersion of $E_k$ is linear ($\frac{dE_w}{dw}=1$) for $w<-\log M\approx -7.6$ or layer $n\approx 14$, but quadratic at momentum above that, till $k\sim O(1)$ where the dispersion levels off.} \label{fig:diracM}\end{figure}

A simple plot reveals that the dispersion $E_k=\sqrt{\sin^2 k + M^2(1-\cos k)^2}$ looks almost perfectly quadratic for $M>5$. However, the short linear region near $k=0$ is still expected to dominate the physics at the IR layers of the bulk. To see that this is indeed true, we explicitly calculate the order of the dispersion which is given by the derivative $\frac{d\log E_w}{dw}$, where $w=\log k$:
\begin{eqnarray}
\frac{d\log E_w}{dw}&=& \frac{e^w \left(-M^2+\left(-1+M^2\right) \cos\left[e^w\right]\right) \cot\left[\frac{e^w}{2}\right]}{-1-M^2+\left(-1+M^2\right) \cos\left[e^w\right]}\notag\\
\end{eqnarray}
For small negative values of $w=\log k$, $\frac{d\log E_w}{dw}=2(1-M^2e^{-2w})=2\left(1-\frac{M^2}{k^2}\right)\approx 2$. But $\frac{d\log E_w}{dw}\rightarrow 1$ for large negative $w$. The transition region occurs at $k_c\approx \frac{1}{M}$, as shown in Fig. \ref{fig:diracM}. Note that a simple Taylor expansion will \emph{not} reveal a quadratic dispersion, because it is unable to concentrate on an exponentially small IR region.

The shape of the dispersion discussed above affects the bulk geometry profoundly at nonzero temperature. The emergent black hole radius behaves like $b\propto T^{\frac{1}{\gamma}}$, when $E_k\propto k^\gamma$. Hence $M$ sets the critical temperature which separates the $b\propto T$ and $b\propto \sqrt{T}$ regimes.

\subsubsection{Imaginary time correlator}

The zero-temperature asymptotics of the imaginary time correlation function are only dependent on the extreme IR behavior of the Hamiltonian. As evident in Eq. \ref{time}, a very large $\tau$ in the imaginary time correlator suppresses contributions from all but the lowest energy regime. As such, a finite $M$ should not change the long time behavior of $C_n(\tau)$.

This is however not true for an infinite $M$, which will produce a purely quadratic dispersion since $k_c\approx \frac{1}{M}=0$. 
Let us write $E_k=v_0k^\gamma$ for small $k$, where $\gamma=2$ here. Then Eq. \ref{time2} becomes
\begin{eqnarray}
\text{Tr } C_{n}(\tau)&\approx&\frac{2^{3n+1}}{32 \pi}
\int_{0}^\pi dq  q^{2}e^{-\tau v_0 q^\gamma}\notag\\
&=&\frac{2^{3n}}{16 \pi} \frac{\Gamma\left[\frac{3}{n}\right]}{\gamma (v_0\tau)^{\frac{3}{\gamma}}}
\end{eqnarray}
Hence a purely nonlinear dispersion of order $\gamma$ affects the long time correlator by changing the exponent in the power-law decay from $3$ to $3/\gamma$.
In our current context with the Dirac model, there exists a moderate imaginary time regime where the correlator decays like $\sim\tau^{-\frac{3}{2}}$. The duration of this regime becomes longer and longer as $M$ increases, till it finally becomes infinitely long at $M=\infty$.

\subsubsection{Spatial correlators and criticality}

In general, the power law decay of the spatial correlators depends on the EHM basis, and cannot be changed unless the critical point becomes degenerate. To be precise, the power law decay depends on the existence of a branch cut in the complexified correlator.

In our two-band case, the complexified correlator $\frac{h_z}{E_z}$, which is introduced in Section IV just before Table I, has nonzero elements given by
\begin{equation}
\sqrt{\frac{\sin k + iM(1-\cos k)}{\sin k - iM(1-\cos k)}}=\sqrt{\frac{i(z-z^{-1})-iM(2-(z+z^{-1}))}{i(z-z^{-1})+iM(2-(z+z^{-1}))}}
\end{equation}
and its reciprocal. They have square-root branch points at $z=1$, $z=\frac{M^2-1}{M^2+1}$ and $\frac{M^2+1}{M^2-1}$. To compute the correlator, we perform the contour integral around $|z|=1$ around the branch cut from $z=1$ to $z=\frac{M^2-1}{M^2+1}$, like what was done in Eq. \ref{u2}. The correlator matrix elements are thus proportional to
\begin{equation}
2\int_{\frac{1-M^2}{1+M^2}}^1 W^*_n(z^{-1})W_n(z)z^{2^n\Delta x}\sqrt{F(z)}\frac{dz}{z}
\end{equation}
where $F(z)=\sqrt{\frac{\left(z-\frac{1-M^2}{1+M^2}\right)}{\left(z-\frac{1+M^2}{1-M^2}\right)}}$, or its reciprocal. Via the same steps leading to Eq. \ref{u3}, we will eventually  find a $\sim\frac{1}{x^3}$ decay in the spatial correlator, \emph{as long as the $z=1$ branch point is present}.

The branch point at $z=1$ may disappear when it combines with another branch point. In our case, it happens when $M=\infty$. Then $F(z)$ becomes trivially equal to unity, and the correlator is identically zero.

There are other more interesting degenerate cases where we end up with a spatial correlator that decays \emph{exponentially}, even though the system is gapless. This happens when $F(z)$ has singularities within the unit circle, while the gapless point on the unit circle is not a branch point. An example is given by
\begin{equation}
H(k)=\sin^2 k \sigma_1 + (1-\cos k )\sigma_2
\end{equation}
whose singularities in $z=e^{ik}$ occur at $|i-1+i\sqrt{1+2i}|=0.346$, $1/0.346$ and $1$, with $1$ being a double root that cancels off in $\frac{h_z}{E_z}$. Hence its correlator decays like $\sim 0.346^{2^nx}$. Physically, the gapless point at $z=1$ is not critical because the two bands touch but do not intersect. The exponential decay arises from the effective mass scale due to the curvature of the dispersion.

For two-band models, gapless points of even order are always noncritical (degenerate). However, such points may be critical if there are more than two bands. In general, an $N$-band gapless point will be noncritical if the order of its dispersion is a multiple of $N$.

\section{Derivation of results for higher-dimensional critical systems at $T=0$}
\subsection{Decay of the imaginary time correlator for general $D$}
\label{app:multidimtime}

It is instructive to first perform the derivation for a $(2+1)$-dim boundary system. From Eq. \ref{time}, we have
\begin{eqnarray}
\text{Tr} C_n(\tau)
&=&\int_{-\pi}^\pi\int_{-\pi}^\pi dq_x dq_y |\tilde W_n(e^{iq_x})|^2 |\tilde W_n(e^{iq_y})|^2e^{- E_{\bold q} \tau}\notag\\
\label{time2d}
\end{eqnarray}
where $\tilde W_n$ is the $(1+1)$-dim  bulk projector and $\text{Tr}$ is a trace over the band indices, \emph{not} the $\upsilon$ indices (suppressed for now) labeling the $2^D-1$ bulk sectors containing various combinations of one-dimensional holographic basis vectors. For large $\tau$, it suffices to consider the contributions close to the IR point $\bold q=0$ where $W_n(e^{iq_j})$ is maximal and $E_{\bold q}\approx \sqrt{q_x^2+q_y^2}$. As explained in the main text, $\tilde W_n(e^{iq_j})$ either behaves like a constant or is linear in $q_j$ near $q_j=0$, depending on whether the IR or UV projector is chosen. In this appendix, we shall derive the forms of \emph{all} the terms in the correlator $\text{Tr} C^{\upsilon_1\upsilon_2}_n$, and not just the dominant terms.

Let us write the $\upsilon$ index in binary form $(\kappa_1,\kappa_2,...,\kappa_D)$, where $\kappa_j=0,1$ depending on whether the leading factor of $W_n^\upsilon(z)$ corresponds to an IR or UV projector. From Eq. \ref{u6} and the definition of the projectors in the main text, we know that $|W_n(e^{iq_j})|^2\approx \zeta_j^2 q_j^{2\kappa_j}$ for $q_j<\frac{2\pi}{2^n}$, where (letting $v_F=1$ for simplicity)
\begin{equation}
\zeta_j^2=\frac{1}{2\pi} 2^{(-1)^{\kappa_j}+(2\kappa_j+1)(n-1)}
\end{equation}
We next perform the integral in Eq. \ref{time2d} iteratively, starting from the integral over $q_x$:
\begin{equation}
\text{Tr} C(\tau)\approx \int_{-\pi}^\pi A_x^2A_y^2 k_y^{2\kappa_y} J_{k_y,\kappa_x}(\tau) dk_y
\label{time2d2}
\end{equation}
where $J_{k_y,\kappa_x}$ is an effective massive $(1+1)$-dim  correlator. For large $\tau> \frac{1}{m}$ and $\kappa_x=1$, it can be approximated by
\begin{eqnarray}
J_{k,\kappa_x=1}(\tau)&=& \int_{-\pi}^\pi  e^{-\sqrt{k^2+q^2}\tau} q^{2\kappa_x}dq\notag\\
&\approx & 2\int_{0}^\infty  e^{-\sqrt{k^2+q^2}\tau} q^{2\kappa_x}dq \notag\\
&= & 2\int_{k}^\infty  e^{-\epsilon\tau} (\epsilon^2-k^2)^{\kappa_x}\frac{\epsilon}{\sqrt{\epsilon^2-k^2}}d\epsilon \notag\\
&= & 2\int_{k}^\infty  e^{-\epsilon\tau} (\epsilon^2-k^2)^{\kappa_x-1}\frac{\epsilon(\epsilon^2-k^2)}{\sqrt{\epsilon^2-k^2}}d\epsilon \notag\\
&\approx  & 2\int_{k}^\infty  e^{-\epsilon\tau} (\epsilon^2-k^2)^{\kappa_x-1}\left(\epsilon^2-\frac{k^2}{2}\right)d\epsilon \notag\\
&=  & 2\int_{k}^\infty  e^{-\epsilon\tau} \left [(\epsilon^2-k^2)^{\kappa_x}+\frac{k^2}{2}(\epsilon^2-k^2)^{\kappa_x-1}\right]d\epsilon \notag\\
&=& e^{-k\tau}\frac{(2+k\tau)^2}{\tau^3}\notag\\
&=& \frac{e^{-k\tau}}{\tau^3}Q_{\kappa_x=1}(\tau k)
\label{time2d3}
\end{eqnarray}
where $Q_{\kappa}$ is a $2\kappa$-th degree polynomial with constant term $(2\kappa)!$. The approximation from line 1 to 2 is extremely accurate for large $\tau$, while that from line 4 to 5 is valid for for extremely small $k$. This is the regime that contributes most to $\text{Tr} C_n(\tau)$, because $J_{k,\kappa_x}$ is suppressed by at least like $e^{-k\tau}$ where $\tau$ is large. For small $k$, the integrand does not decay fast, and indeed it is the regime where $u\gg k$ that contributes most to the integral. 
The other case, $J_{k,\kappa_x=0}(\tau)$, resist all known approaches of analytical approximation. However, it is obvious that it behaves asymptotically like
\begin{equation}  J_{k,\kappa_x=0}(\tau)\sim \frac{e^{-k\tau}}{\tau} \end{equation}
from the relation $(\partial_\tau^2-k^2)J_{k,\kappa_x=0}(\tau)=J_{k,\kappa_x=1}(\tau)$, i.e. with $Q_{\kappa_x=0}(\tau k)\sim \text{const}$.

For a critical system in $D+1$ dimensions, we just have to replace
$E_q$ by $v_F\sqrt{\sum_{j=1}^{D} q_j^2}$, and substitute that into Eq. \ref{time2d}. Now let's define $T_j(\tau)$ to be the integrand of $\text{Tr} C_n(\tau)$ with the first $j$ dimensions integrated over. Our goal is to find the asymptotic behavior of $T_D(\tau)$. We have
\begin{equation}
T_i(\tau)\approx e^{-P_i \tau }\left[\prod^i_{j=1}\int_{-\pi}^{\pi}dq_j \zeta^2_j q_j^{2\kappa_j}\right]
\end{equation}
where $P_i=\sqrt{\sum_{j=1}^i q_j^2}$. We perform the integral over last variable using the same approximations (valid for large $\tau$) as in Eq. \ref{time2d3}:
\begin{eqnarray}
T_i(\tau)&\approx &\left[\prod^{i-1}_{j=1}\int_{-\pi}^{\pi}dq_j \zeta^2_j q_j^{2\kappa_j}\right]\int_{-\pi}^{\pi}dq \zeta^2_i q^{2\kappa_i}e^{-P_i \tau  }\notag\\
&= &2\zeta^2_i\left[\prod^{i-1}_{j=1}\int_{-\pi}^{\pi}dq_j \zeta^2_j q_j^{2\kappa_j}\right]\int_{0}^{\infty}dq  q^{2\kappa_i}e^{-\sqrt{P_{i-1}^2+q^2} \tau  }\notag\\
&\approx  &2\zeta^2_i\left[\prod^{i-1}_{j=1}\int_{-\pi}^{\pi}dq_j \zeta^2_j q_j^{2\kappa_j}\right]Q_{\kappa_i}(\tau P_{i-1} )\frac{e^{-P_{i-1} \tau  }}{\tau^{2\kappa_i+1}}\notag\\
&= &2\zeta^2_i\tilde Q_{\kappa_i}\left(-\tau \frac{d}{d\tau} \right)\left(\frac{1}{\tau^{2\kappa_i+1}}T_{i-1}(\tau)\right)\notag\\
&= &2^{i-1}\left[\prod_{j=2}^{i}\zeta^2_j\tilde Q_{\kappa_j}\left(-\tau \frac{d}{d\tau} \right)\right]\left(\frac{1}{\tau^{2(\sum_{j=2}^i\kappa_j)+i-1}}T_{1}(\tau)\right)\notag\\
&= &2^i (2\kappa_1)!\prod_{j=1}^{i}\zeta^2_j\left[\prod_{j=2}^{i}\tilde Q_{\kappa_j}\left(-\tau \frac{d}{d\tau} \right)\right]\frac{1}{\tau^{2(\sum_{j=1}^i\kappa_j)+i}}\notag\\
&\propto &\prod_{j=1}^{i}\zeta^2_j\frac{1}{\tau^{2(\sum_{j=1}^i\kappa_j)+i}}\notag\\
&\sim& \prod_{j=1}^{i}\left(\frac{2^{n-1}}{\tau}\right)^{2\kappa_i+1}
\label{timeD}
\end{eqnarray}
On line 4, the tilde in $\tilde Q_{\kappa}\left(-\tau \frac{d}{d\tau} \right)$ denote normal ordering of the $\tau$ and $\frac{d}{d\tau}$ operators, i.e. $Q_{\kappa_j}$ and its products will have all $\tau$ moved to the left of all $\frac{d}{d\tau}$. When going from the sixth to seventh line, we note that each operator $(-\tau^n)\frac{d^n}{d\tau^n}\rightarrow \frac{(n+\kappa)!}{\kappa !}$ when acting on expressions of the form $1/\tau^{\kappa+1}$, without incurring additional factors of $1/\tau$.

Hence
\begin{equation}
\text{Tr} C_n(\tau)\sim \frac{1}{v_F^D\tau^D}\prod_{j=1}^D \frac{2^{(2\kappa_j+1)(n-1)}}{(v_F\tau)^{2\kappa_j}}
\label{generaltau}
\end{equation}
, after restoring $v_F$. Evidently, the leading terms occur when $\kappa_j=1$ for just one $j$, and are zero for the others. Hence, we have
\begin{equation}
\text{Tr } C_n(\tau)\sim \left(\frac{2^n}{v_F \tau}\right)^{D+2}
\label{generaltau2}
\end{equation}
This is the main result for the imaginary time correlator in the multidimensional critical case.

For the multidimensional massive case with a fixed mass $m$, $P_i$ in Eq. \ref{timeD} is replaced by $P_i=\sqrt{m^2+\sum_{j=1}^i q_j^2}$. Hence $T_1(\tau)$ also contains a mass and the third last line of Eq. \ref{timeD} becomes \[2^i \left[\prod_{j=1}^{i}\zeta^2_j\tilde Q_{\kappa_j}\left(-\tau \frac{d}{d\tau} \right)\right]\frac{e^{-m\tau}}{\tau^{2(\sum_{j=1}^i\kappa_j)+i}},\]
valid for $m\tau<1$ due to the approximations in Eq. \ref{time2d3}. Hence

Eq. \ref{generaltau} becomes
\begin{eqnarray}
&&\text{Tr} C_n(\tau)|_{0<m\tau<1}\notag\\
&\sim& \frac{e^{-m\tau}}{v_F^2\tau^2}\prod_{j=1}^D \frac{2^{(2\kappa_j+1)(n-1)}}{(v_F\tau)^{2\kappa_j}} + \text{higher orders of $1/\tau$}\notag\\
\label{generaltaum}
\end{eqnarray}
and
\begin{equation}
\text{Tr} C_n(\tau)|_{m\tau\gg 1}\sim e^{-m\tau} \times \text{weak dependence on powers of  $1/\tau$}\notag\\
\label{generaltaum2}
\end{equation}
The nonzero  mass correlator is exponentially suppressed by $e^{-m\tau}$, though for small $m\tau$ we still see a subleading power law in $\tau$, albeit with a different power from that of the massless case.

\subsection{Nonuniversal properties of $(2+1)$-dim critical model}
\label{app:nonuniversal2d}

Here we illustrate how a different choice of model in $(2+1)$-dim can affect certain quantities but not others. We consider the model
\begin{eqnarray}
H(q_x,q_y)&= &d(q)\cdot \sigma\notag\\
&=&\sin q_x \sigma_1+\sin q_y \sigma_2 +(\cos q_x-\cos q_y)\sigma_3\notag\\
\end{eqnarray}
which is also gapless at $q=0$. Its correlator in momentum space is
\begin{eqnarray}
G_q&=&\frac{1}{2}\left( \mathbb{I}-\hat d(q)\cdot \sigma \right)\notag\\
&=&\frac{1}{2}\left(\begin{matrix}
 & 1 -\frac{\cos q_x -\cos q_y}{E_q}& -\frac{\sin q_x -i\sin q_y}{E_q} \\
 & -\frac{\sin q_x +i\sin q_y}{E_q} & 1+\frac{\cos q_x -\cos q_y}{E_q} \\
\end{matrix}\right)
\label{gq2d}
\end{eqnarray}
where $E_q= \sqrt{2(1-\cos q_x \cos q_y)}\rightarrow \sqrt{q_x^2+q_y^2}=|q|$ near criticality. When considered as a $(1+1)$-dim  correlator depending on $q_x$($q_y$) alone, it behaves like a massive correlator with mass $q_y$($q_x$), as can be seen from its poles at $\pm i\cosh^{-1}\sec q_y$ (and vice versa for $q_x\leftrightarrow q_y$).

Since the single-site correlator $C_x$ is given by
\begin{equation}C_x=\int_{-\pi}^\pi\int_{-\pi}^\pi dq_x dq_y |W_n(e^{iq_x})|^2 | W_n(e^{iq_y})|^2 G_{\bold q}\end{equation}
where $ |W_n(e^{iq_j})|^2$ is even about $q=0$, we see that the off-diagonal components, being odd in $q_x$ or $q_y$, must disappear. This is different from the $1+1$-dim Dirac model, where $d_2(q)$, which is not odd in $q$,  plays an important role in the decay of the correlator. In the current $(2+1)$-dim case, it is $d_3$ that controls the decay of the correlator.

Juch as importantly, note that the nonconstant part of the diagonal terms are odd under the interchange $q_x\leftrightarrow q_y$. Hence $G_q\propto \frac{1}{2}\mathbb{I}$ due to the symmetry betwee the wavelet bases $W_n(e^{iq_x})$ and $W_n(e^{iq_y})$. With the off-diagonal part $A$ (defined previously) vanishing rigorously, the single-site entropy is always maintained at exactly $S_x=\log 4$, the same universal limiting value in the $(1+1)$-dim  case. Evidently, the small $n$ (UV) behavior of the entropy depends nonuniversally on the details of the model.  





\section{Relationship between real and imaginary time correlators}
\label{sec:time}

Our discussion of the EHM will not be complete without a proper discussion of the real time correlator, which is arguably of more direct physical significance. However, its oscillatory nature makes it unsuitable as a definition of bulk distance. Here, we shall discuss its mathematical and physical significance with the imaginary time correlator.
\subsection{Critical case with linear dispersion}

When there is a linearly dispersive critical point $E_q = v_F q$, Galilean invariance is restored and there is symmetry between space and time. Restricting ourselves again to $(1+1)$-dimensions, the bulk correlator within a layer is explicitly given by
\begin{eqnarray}
C(n,n,\Delta x, \Delta t)&=&\sum_q  |W_{n}(q)|^2 e^{i2^nq\Delta x}G_q(-i\Delta t)\notag\\
&=&\sum_q   |W_{n}(q)|^2 e^{i2^nq\Delta x}e^{i E_q \Delta t}G_q\notag\\
&=& \sum_q  |W_{n}(q)|^2 e^{iq[2^n\Delta x+ v_F \Delta t ]}G_q\notag\\
\label{realtime1}
\end{eqnarray}
which depends symmetrically on $2^n \Delta x$ and $v_F \Delta t$. From Eq. \ref{u4}, we deduce that the magnitude of the real time correlator behaves like
\begin{equation}
|C(n,-i\Delta t)|=u|_{\Delta x =2^{-n}v_F \Delta t}\sim \frac{1}{16\pi}\left(\frac{2^n}{v_F \Delta t}\right)^3
\end{equation}
which is identical to that of the imaginary time correlator\footnote{Its \emph{phase}, however, exhibits a nonuniversal and generally non-oscillatory behavior that depends on the details of the model}. This conclusion is consistent with the result obtained by a naive Wick rotation, since extra factors of $i$ in a power-law do not affect the decay behavior.

When the dispersion is nonlinear, Galilean invariance is lost and the correct result cannot be simply obtained via Wick rotation, even if the system is still critical.

\subsection{Non-critical cases}

When a mass scale is present, the energy $E_q$ is bounded below by a value $m$, i.e. $E_q=m+ \epsilon_q$ where $\epsilon_q\geq 0$. Hence, the real time bulk correlator
\begin{eqnarray}
C(n_1,n_2,0, \Delta t)
&=&\sum_q  W_{n_1}^*(q)W_{n_2}(q) e^{i E_q \Delta t}G_q\notag\\
&=&e^{im\Delta t}\sum_q  W_{n_1}^*(q)W_{n_2}(q) e^{i \epsilon_q \Delta t}G_q\notag\\
\label{realtime2}
\end{eqnarray}
acquires an oscillatory phase with frequency $m$, a result again consistent with Wick-rotating the exponential decay $e^{-m\tau}$ behavior of the imaginary time correlator.

\section{The behavior of geodesic distance}
\label{app:geodesics}
Some results of this subsection and the next can also be found in Ref. \onlinecite{qi2013}. Here, we reproduce them for completeness.

\subsection{AdS space}

Due to its remarkable symmetry, Anti-de-Sitter space can be embedded in a flat Minkowski spacetime one dimension higher. As such, it inherits the simple metric structure of the latter, which yields simple expressions for geodesic distances.

For brevity, lets only consider the Euclidean $(2+1)$-dim AdS space, since its higher dimensional analogues will give rise to very similar expressions. We parametrize the space by coordinates $w=(\rho,\theta,\tau)$, and embed it in a 4-d Minkowski spacetime as the locus of $X^ax^b\eta_{ab}=X\cdot X = R^2$, where $R$ is the AdS radius, $\eta=diag(1,-1,-1,-1)$ and
\begin{widetext}
\begin{eqnarray}
X^\mu(w)&=& \left(\sqrt{\rho^2+R^2}\cosh \frac{ \tau}{\sqrt{R^2+\frac{L^2}{4\pi^2}}},\sqrt{\rho^2+R^2}\sinh\frac{ \tau}{\sqrt{R^2+\frac{L^2}{4\pi^2}}},\rho \cos \theta, \rho \sin \theta\right)
\label{geod1}
\end{eqnarray}
\end{widetext}
Here $\tau$ appears with a rescaling factor of $\frac{1}{\sqrt{R^2+\frac{L^2}{4\pi^2}}}$, instead of the more conventional $\frac{1}{R}$. This is to ensure that the rescaled AdS metric
\begin{equation}
\frac{1+\frac{\rho^2}{R^2}}{1+\frac{L^2}{4\pi^2R^2}}d\tau^2 + \frac{d\rho^2}{1+\frac{\rho^2}{R^2}}+\rho^2 d\theta^2\rightarrow d\tau^2 + \rho^2 d\theta^2
\label{adsmetric}
\end{equation}
is $O(2)$-invariant at $\rho=\frac{L}{2\pi}\gg R$.
From well-known properties of the Minkowski metric, the geodesic distance between points $w_1$ and $w_2$ is given by
\begin{equation}
d_{12}=R\cosh^{-1}\left(\frac{X(w_1)\cdot X(w_2)}{R^2}\right)
\label{geod2}
\end{equation}
From Eq. \ref{geod2}, we find the geodesic distance corresponding to an angular displacement 
to be
\begin{eqnarray}
d^{min}_{\Delta \theta}&=&R\cosh^{-1}\left(\frac{\rho^2 + R^2 - \rho^2 \cos \Delta \theta}{R^2}\right)\notag\\
&=& R\cosh^{-1}\left(1+\frac{2\rho^2}{R^2}\sin^2\frac{\Delta \theta}{2}\right)\notag\\
&\sim & 2R\log \frac{\rho \sin\Delta \theta}{R}\notag\\
&\approx & 2R\log \frac{\rho \Delta \theta}{R}
\label{ads1}
\end{eqnarray}
for $\rho \gg R$. In the scale-invariant case where $\Delta x =\rho \Delta \theta$, and we also have $d^{min}_{\Delta x}\sim 2R \log \frac{|\Delta x|}{R}$.

For a radial displacement with $\Delta \rho = |\rho_2-\rho_1|$, we have
\begin{eqnarray}
d^{min}_{\Delta \rho}&=&R\cosh^{-1}\left(\frac{\sqrt{(R^2+\rho_1^2)(R^2+\rho_2^2)}-\rho_1\rho_2}{R^2}\right)\notag\\
&\approx &R\cosh^{-1}\left(\frac{\rho_1}{R}\frac{\rho_2}{R}\left(\left(1+\frac{R^2}{2 \rho_1^2}\right)\left(1+\frac{R^2}{2 \rho_2^2}\right)-1\right)\right)\notag\\
&\approx& R\cosh^{-1}\left(\frac{1}{2}\left(\frac{\rho_1}{\rho_2}+\frac{\rho_2}{\rho_1}\right)\right)\notag\\
&=& R \left|\log\frac{\rho_1}{\rho_2}\right|\notag\\
&=& R|\Delta(\log \rho)|,
\label{ads2}
\end{eqnarray}
also for $\rho_1,\rho_2 \gg R$.

The geodesic distance in the imaginary time direction is given by
\begin{eqnarray}
d_\tau &=&R\cosh^{-1}\left(\left(\frac{\rho^2}{R^2}+1\right)\cosh\frac{2\pi \tau}{L^2 + 4\pi^2 R^2}-\frac{\rho^2}{R^2}\right)\notag\\
&\approx &R\cosh^{-1}\left(\frac{\rho^2}{R^2}\left(\cosh\frac{2\pi \tau}{L^2 + 4\pi^2 R^2}-1\right)\right)\notag\\
&\approx& R\cosh^{-1}\left(\frac{\rho^2}{R^2}\frac{(2\pi)^2 \tau^2}{2(L^2 + 4\pi^2R^2)}\right)\notag\\
&\sim& 2R\log \frac{2\pi \rho \tau}{RL}
\label{ads3}
\end{eqnarray}
where we have used $\rho\gg R$ while going from line $1$ to $2$, and $t\ll L$ from line $2$ to $3$ and $\rho \tau \gg RL$ from line $3$ to $4$.

\subsection{Geodesics near a black hole horizon}
\label{app:rindler}

We start from a metric
\begin{equation}
ds^2= V(r)d\tau^2 + \frac{dr^2}{V(r)}+ r^2 d\Omega^2 
\end{equation}

with $V(r_0)=0$, which admits a horizon at $r=r_0$. This horizon is a null surface within which $ds^2=0$. Intuitively, the geodesics between two points infinitesimally close to the horizon must necessarily wrap around the horizon (i.e. with with $r$ being constant), since the cost of radial displacements $\Delta r $ diverges to infinity at the horizon. Hence we always have $\Delta s \approx r_0 \Delta \theta$ near a horizon.

To explore the near-horizon geometry more rigorously, we switch to Rindler coordinates valid near the horizon $r=r_0$. We define
\begin{equation}
\rho = 2 \text{sgn}(r-r_0)\sqrt{\frac{|r-r_0|}{|V'(r_0)|}}
\end{equation}
so that, to order $O(\rho^2)$, the metric becomes
\begin{widetext}
\begin{eqnarray}
ds^2|_R&=& \rho^2\left(1+\frac{V''(r_0)}{8}\rho^2\right)dT^2 + \left(1-\frac{V''(r_0)}{8}\rho^2\right)d\rho^2 + \left(r_0\pm \frac{|V'(r_0)|\rho^2}{4}\right)^2 d\Omega^2
\end{eqnarray}
\end{widetext}
where $T=\frac{V'(r_0)}{2}\tau$ and the $\pm$ sign refers to the region immediately outside and inside the horizon respectively. 

Upon dropping the 2nd order terms, we recover the usual Rindler metric which just describes a plane with $T$ taking the role of the polar angle. To avoid a conical singularity, we require that $T$ has period $2\pi$, i.e. that the period of $\tau$ and thus $\beta= T^{-1}$ is of the value $\frac{4\pi}{V'(r_0)}$.

We now specialize to an example most relevant to the main text, which is the near-horizon geometry of a $(2+1)$-dim BTZ black hole. The horizon occurs at $r=b$, with another parameter $R$ setting the overall length scale. The horizon must also occur at a $D+1$-th order zero of $V(r)$. Hence we have
\begin{equation}
V(r)=\frac{r^3-b^3}{R^2r}
\end{equation}
with $V'(b)=\frac{3b}{R^2}$ and $V''(b)=0$. That the second derivative is identically zero is unique to $D=2$ dimensions. After some tedious derivation, the geodesic distance at Rindler radius $\rho_0$ between two points separated by $\Delta \theta$ is
\begin{eqnarray}
&&\Delta \lambda\notag\\ &=&\sqrt{\frac{2}{3}}\frac{R}{\sqrt{1+\eta^2}}\cosh^{-1}\left[ \sqrt{1+\eta^2}\cosh \left(\sqrt{\frac{3}{2}} \frac{\sqrt{1+\eta^2}}{R}b\Delta \theta\right)\right]\notag\\
\end{eqnarray}
where $b$ is the horizon radius and
\begin{equation} \eta=\sqrt{\frac{3}{2}}\frac{\rho_0}{R}\text{ sech}\sqrt{\frac{3}{2}}\frac{b\Delta\theta}{R}
\end{equation}
A moment of calculation reveals that the above reduces to $\Delta\lambda = b\Delta\theta$ as we approach the horizon where $\eta\propto \rho_0\rightarrow 0$.



\end{document}